\documentclass[aps,prd,showpacs,nofootinbib,sort&compress]{revtex4-1}
\usepackage{blindtext}
\usepackage{amsmath}
\usepackage{amssymb}
\usepackage{framed}
\expandafter\ifx\csname package@font\endcsname\relax\else
\expandafter\expandafter
 \expandafter\usepackage
 \expandafter\expandafter
 \expandafter{\csname package@font\endcsname}%
\fi
\usepackage[utf8]{inputenc}
\usepackage{bm}
\usepackage{url}
\usepackage{color}
\usepackage{times}
\usepackage{graphicx}
\usepackage{textcomp}
\usepackage[update,prepend]{epstopdf}
\usepackage{rotating}
\usepackage{txfonts} 
\providecommand{\url}[1]{{#1}}

\expandafter\ifx\csname urlstyle\endcsname\relax
  \else
  \fi

\def\be{\begin{equation}}
\def\ee{\end{equation}}
\def\ba{\begin{eqnarray}}
\def\ea{\end{eqnarray}}

\def\a{\alpha}
\def\a {\alpha}
\def\b{\beta}
\def\g{\gamma}     
\def\d{\delta}     \def\D{\Delta}

\def\e{\epsilon}

\def\l{\lambda}

\def\o{\omega}

\def\th{\theta}

\def\r{\rho}
\def\s{\sigma}

\def\la{\label}
\def\pd{\partial}
\def\lt{\left}
\def\rt{\right}

\begin{document}
\title{Post-Newtonian reference-ellipsoid for relativistic geodesy.}
\author{Sergei Kopeikin}
\email{kopeikins@missouri.edu}
\affiliation{Department of Physics \& Astronomy, 322 Physics Bldg, University of Missouri, Columbia, Missouri
65211, USA}
\altaffiliation[Also at: ]{Siberian State University of Geosystems and Technologies, Plakhotny St. 10,
Novosibirsk 630108, Russia}
\author{Wenbiao Han}
\email{wbhan@shao.ac.cn}
\affiliation{Shanghai Astronomical Observatory, 80 Nandan Road, Shanghai, 200030, China}
\author{Elena Mazurova}
\email{e\_mazurova@mail.ru}
\affiliation{Siberian State University of Geosystems and Technologies, Plakhotny St. 10, Novosibirsk 630108,
Russia}
\altaffiliation[Also at: ]{Moscow State University of Geodesy and Cartography, 4 Gorokhovsky Alley, Moscow
105064,\phantom{Moscow State University of Geodesy and Cartography, 4 Gorokhovsky Alley, Moscow
105064}
Visiting Scholar:  Department of Physics \& Astronomy, 223 Physics Bldg, University of Missouri, Columbia, Missouri
65211, USA}

\date{\today}
\begin{abstract}
We apply general relativity to construct the post-Newtonian background manifold that serves as a reference spacetime in relativistic geodesy for conducting relativistic calculation of the geoid's undulation and the deflection of the plumb line from the vertical. We chose an axisymmetric ellipsoidal body made up of perfect homogeneous fluid uniformly rotating around a fixed axis, as a source generating the reference geometry of the background manifold through Einstein's equations. We, then, reformulate and extend hydrodynamic calculations of rotating fluids done by a number of previous researchers for astrophysical applications to the realm of relativistic geodesy to set up algebraic equations defining the shape of the post-Newtonian reference ellipsoid. To complete this task, we explicitly perform all integrals characterizing gravitational field potentials inside the fluid body and represent them in terms of the elementary functions depending on the eccentricity of the ellipsoid. We fully explore the coordinate (gauge) freedom of the equations describing the post-Newtonian ellipsoid and demonstrate that the fractional deviation of the post-Newtonian level surface from the Maclaurin ellipsoid can be made much smaller than the previously anticipated estimate based on the astrophysical application of the coordinate gauge advocated by Bardeen and Chandrasekhar. We also derive the gauge-invariant relations of the post-Newtonian mass and the constant angular velocity of the rotating fluid with the parameters characterizing the shape of the post-Newtonian ellipsoid including its eccentricity, a semiminor and a semimajor axes. We formulate the post-Newtonian theorems of Pizzetti and Clairaut that are used in geodesy to connect the geometric parameters of the reference ellipsoid to the physically measurable force of gravity at the pole and equator of the ellipsoid. Finally, we expand the post-Newtonian geodetic equations describing the post-Newtonian ellipsoid to the Taylor series with respect to the eccentricity of the ellipsoid and discuss their practical applications for geodetic constants and relations adopted in fundamental astronomy.
\end{abstract}
\maketitle

\section{Introduction}\la{sec1}

Accurate definition, determination and realization of celestial and terrestrial reference frames in the solar
system is essential for deeper understanding of the underlying principles and concepts of fundamental
gravitational physics, astronomy and geophysics as well as for practical purposes of precise satellite and
aircraft navigation, positioning and mapping. It was suggested long ago to separate the conceptual meaning of a
reference frame and a reference system \citep{koval_1989}. The former is understood as a theoretical
construction, including mathematical models and standards for its implementation. The latter is its practical
realization through observations and materialization of coordinates of a set of reference benchmarks, e.g., a
set of fundamental stars - for the International Celestial Reference Frame (ICRF) - or a set of fiducial
geodetic stations - for the International Terrestrial Reference Frame (ITRF). Continuous monitoring and maintenance of the reference frames and a self-consistent set of geodetic and astronomical constants associated with them, is rendered by the International Earth Rotation Service (IERS) and the International Union of Geodesy and Geophysics (IUGG). 

Nowadays, four main geodetic techniques are used to compute accurate terrestrial coordinates and velocities of
stations -- GPS, VLBI, SLR, and DORIS, for the realizations of ITRF referred to different epochs. The
observations are so accurate that geodesists have to model and to include to the data processing the secular
Earth's crust changes to reach self-consistency between various ITRF realizations which are available on
\textcolor{blue}{\url{http://itrf.ensg.ign.fr/ITRF_solutions/index.php}}. The higher frequencies of the station
displacements (mainly due to geophysical phenomena) can be accessed with the formulations present in chapter 7
of IERS conventions \citep{petit_2010} (see also
\textcolor{blue}{\url{http://62.161.69.131/iers/convupdt/convupdt.html}}). Continuity between the ITRF
realizations has been ensured when adopting conventions for ICRF and ITRF definitions \citep[chapter
4]{petit_2010}. It is recognized that to maintain the up-to-date ITRF realization as accurate as possible the
development of the most precise theoretical models and parametric relationships is of a paramount importance.

Currently, SLR and GPS allow us to determine the transformation parameters between coordinates and velocities
of the collocation points of the ITRF realizations with the precision of $\sim$1 mm and $\sim$1 mm/yr
respectively \citep[table 4.1]{petit_2010}. On the other hand, the dimensional analysis applied to estimate the
relativistic effects in geodesy predicts that the post-Newtonian contribution to the coordinates of the ITRF
points on the Earth's surface (as compared with the Newtonian theory of gravity) is expected to be of the order
of the Earth's gravitational radius that is about 9 mm \citep{Kopejkin_1991,Mueller_2008JGeod} or, may be, less
\citep{will_2014LRR}. This post-Newtonian geodetic effect emerges as an irremovable long-wave deformation of the
three-dimensional coordinate grid of ITRF which might be potentially measurable with the currently available
geodetic techniques and, hence, deserves to be taken into account when building the ITRF realization of a next
generation.

ITRF solutions are specified by the Cartesian equatorial coordinates $x^i=\{x,y,z\}$ of the reference geodetic
stations. For the purposes of geodesy and gravimetry the Cartesian coordinates are often converted to
geographical coordinates $h,\th,\lambda$ ($h$ - height, $\th$ - latitude, $\lambda$ - longitude) referred to an international reference ellipsoid which is a
solution found by Maclaurin \citep{Chandr_1967a} for the figure of a fluid body with a homogeneous mass density
that slowly rotates around a fixed $z$-axis with a constant angular velocity $\o$. For the post-Newtonian effects
deforms the shape of the reference-ellipsoid \citep{Chandr_1965ApJ142_1513,Bardeen_1971ApJ} and modify the basic
equations of classic geodesy \citep{Kopeikin_2011_book,Mai_2014}, they must be properly calculated to ensure
the adequacy of the geodetic coordinate transformations at the millimeter level of accuracy. In order to
evaluate more precisely the post-Newtonian effects in the shape of the Earth's reference ellipsoid and the
geodetic equations we have decided to conduct more precise mathematical study of equations of relativistic
geodesy which is given in the present paper. 

Certainly, we are touching upon the topic which has been already discussed in literature by research teams from USA
\citep{Chandr_1965ApJ142_1513,Chandr_1967ApJ147_334,Chandr_1967ApJ148_621,Chandr_1967ApJ148_645,
Chandr_1971ApJ167_447,Chandr_1971ApJ167_455,Chandr_1974MNRAS167,Chandr1974ApJ192_731,Chandr1978ApJ220_303,
Bardeen_1971ApJ}, Lithuania \citep{Pyragas1974Ap&SS27_437,Pyragas1974Ap&SS27_453,Pyragas1975Ap&SS33_75}, USSR \citep{Tsirulev_1982SvA_289,Tsirulev_1982SvA_407,Tsvetkov_1983SvA,Galtsov_1984JETP} and, the most recently,
by theorists from the University of Jena in Germany
\citep{Petroff_2003PhRvD,Ansorg2004MNRAS,Meinel_2008,Gurlebeck2010ApJ722,Gurlebeck2013ApJ777}. We draw attention of the reader that the previous papers focused primarily on studying the astrophysical aspects of the problem like the instability of the equilibrium rotating configurations and the points of bifurcations, finding exact solutions of Einstein's equations for axially-symmetric spacetimes, emission of gravitational waves, etc. Our treatment concerns different aspects and is focused on the post-Newtonian effects in physical geodesy. More specifically, we extend the research on the figures of equilibrium into the realm of relativistic geodesy and pay attention mostly to the possible geodetic applications for an adequate numerical processing of the high-precise data obtained by various geodetic techniques that include but not limited to SLR, LLR, VLBI, DORIS and GNSS \citep{Pearlman_2009book,Drewes_2009book}. This vitally important branch of the theory of equilibrium of rotating bodies was not covered by the above-referenced astrophysical works.

Additional stimulating factor for pursuing more advanced research on relativistic geodesy and the Earth figure of equilibrium is related to the
recent breakthrough in manufacturing quantum clocks \citep{Poli_2013NCimR}, ultra-precise time-scale dissemination
over the globe \citep{kessler_2014NatPh}, and geophysical applications of the clocks
\citep{Bondarescu_2012GeoJI,2015arXiv150602457B}. Clocks at rest in a gravitational potential tick slower than
clocks outside of it. On Earth, this translates to a relative frequency change of $10^{-16}$ per meter of
height difference \citep{Falke_2013}. Comparing the frequency of a probe clock with a reference clock provides
a direct measure of the gravity potential difference between the two clocks. This novel technique has been
dubbed {\it chronometric levelling}. It is envisioned as one of the most promising application of the
relativistic geodesy in a near future \citep{Mai_2013ZGGL,Mai_2014,Petit_2014}. Optical frequency standards
have recently reached stability of $2.2\times 10^{-16}$ at 1 s, and demonstrated an overall fractional
frequency uncertainty of $2.1\times 10^{-18}$ \citep{clock_2015NatCo} which enables their use for relativistic
geodesy at an absolute level of one centimetre.

The chronometric levelling directly measures the equipotential surface of gravity field (geoid) without
conducting a complicated gravimetric survey and solving the differential equations for anomalous gravity
potentials \citep{Petit_2014}. Combining the data of the chronometric levelling with those of the conventional geodetic techniques
will allow us to determine the normal heights of reference points with an unprecedented accuracy
\citep{Bondarescu_2012GeoJI}. An adequate physical interpretation of this type of measurements is inconceivable
without an accompanying development of the corresponding mathematical algorithms accounting for the major
relativistic effects in geodesy.

Basic theoretical concepts of relativistic geodesy have been discussed in a number of textbooks, most notably
\citep{Soffel_1989_book,will_1993book,Kopeikin_2011_book,Torge_2012_book} and review papers \citep{Mai_2014,Petit_2014,will_2014LRR}. Nonetheless,
theoretical problem of the determination of the reference level surface in relativistic geodesy has not yet  been
discussed in scientific literature with a full mathematical rigour. The objective of the present paper is to give
its comprehensive post-Newtonian solution.
To this end, section \ref{sec2} explains briefly the principles of the post-Newtonian approximations and describes
the post-Newtonian metric tensor. Section \ref{sec3} discusses the post-Newtonian ellipsoid which generalizes
the Maclaurin ellipsoid and is the surface of the fourth order. Section \ref{sec4} introduces the reader to the
concept of the post-Newtonian gauge freedom and shows how this freedom can be used to simplify the mathematical
description of the PN ellipsoid. Sections \ref{sec5} and \ref{sec6} calculate respectively the Newtonian and
post-Newtonian gravitational potentials inside the rotating PN ellipsoid. Section \ref{sec7} gives the
post-Newtonian definitions of the conserved mass and angular momentum of the rotating PN ellipsoid. Section
\ref{sec8} is devoted to the derivation of the post-Newtonian equations defining the geometric structure of the reference
level surface and its kinematic relation to the angular velocity of rotation $\o$. Sections \ref{sec9} and \ref{sec10} provide the reader with the relativistic generalization of
the Pizzetti and Clairaut theorem of classical geodesy \citep{pizz_1913,Zund_1994_book} which connect parameters of the reference ellipsoid with the measured value of the force of gravity. Finally, section \ref{sec11} gives truncated versions
of the relativistic formulas which can be used in practical applications of relativistic geodesy. Appendix
\ref{appen1} contains details of the mathematical calculation of integrals.

We adopt the following notations: 
\begin{itemize}
\item the Greek indices $\alpha, \beta,...$ run from 0 to 3,
\item the Roman indices $i,j,...$ run from 1 to 3, 
\item repeated Greek indices mean Einstein's summation from 0 to 3, 
\item repeated Roman indices mean Einstein's summation from 1 to 3,
\item the unit matrix (also known as the Kroneker symbol) is denoted by $\delta_{ij}=\d^{ij}$,
\item the fully antisymmetric symbol Levi-Civita is denoted as $\varepsilon_{ijk}=\varepsilon^{ijk}$ with
$\varepsilon_{123}=+1$,
\item the bold letters ${\bm
a}=(a^1,a^2,a^3)\equiv(a^i), {\bm b}=(b^1,b^2,b^3)\equiv (b^i),$ and so on, denote spatial
3-dimensional vectors, 
\item a dot between two spatial vectors, for example
${\bm a}\cdot{\bm b}=a^1b^1+a^2b^2+a^3b^3=\d_{ij}a^i b^j$, means the Euclidean dot
product, 
\item the cross between two vectors, for example $({\bm a}\times{\bm b})^i\equiv \varepsilon^{ijk}a^jb^k$,
means the Euclidean cross product,
\item we use a shorthand notation for partial derivatives $\partial_\alpha
=\partial/\partial x^\alpha$,
\item $g_{\a\b}$ is the spacetime metric,
\item the Greek indices are raised and lowered with
the metric $g_{\alpha\beta}$,
\item  the Minkowski (flat) space-time metric
$\eta_{\alpha\beta}={\rm diag}(-1,+1,+1,+1)$, it is used to rise and lower indices of the gravitational metric perturbation,
$h_{\alpha\beta}$.
\item $G$ is the universal gravitational constant,
\item $c$ is the speed of light in vacuum,
\item $\o$ is a constant rotational velocity of the Earth,
\item $\rho$ is a constant density of reference-ellipsoid,
\item $a$ is a semimajor axis of the Maclaurin ellipsoid,
\item $b$ is a semiminor axis of the Maclaurin ellipsoid,
\item $\kappa\equiv \pi G\r a^2/c^2$ is a dimensional parameter characterizing the strength of gravitational field,
\item $R_\oplus$ is the mean (volumetric) radius of the Earth, $R_\oplus\simeq 6.3710\times 10^8$ cm,
\item $a_\oplus$ is the equatorial radius of the Earth reference ellipsoid, $a_\oplus\simeq 6.3781\times 10^8$ cm,
\item $b_\oplus$ is the polar radius of the Earth reference ellipsoid, $b_\oplus\simeq 6.3568\times 10^8$ cm.
\end{itemize}
Other notations are explained in the text as they appear.

\section{Post-Newtonian Metric}\la{sec2}

Einstein's field equations represent a system of ten non-linear differential equations in partial
derivatives for the metric tensor, $g_{\a\b}$, and we have to find their solutions for the case of an
isolated rotating fluid. Because the equations are difficult to solve exactly due to their non-linearity, we
apply the post-Newtonian approximations (PNA) for their solution \citep{Chandr_1965ApJ142_1513}.

The PNA are applied in case of slowly-moving matter having a weak gravitational field. This is exactly the situation in the solar system which makes PNA highly appropriate for constructing a relativistic theory of 
reference frame\index{reference frame}s \citep{Soffel_2003AJ} and relativistic celestial mechanics\index{celestial
mechanics!relativistic} in the solar system \citep{Soffel_1989_book,brumberg_1991_book,Kopeikin_2011_book}.

The PNA are based on the assumption that a Taylor expansion of the metric tensor can be done in inverse powers of the fundamental speed $c$ that is equal to the speed of
light in vacuum. Exact mathematical formulation of a set of basic axioms required for doing the
post-Newtonian expansion was given by Rendall \citep{rendall}. Practically, it requires to have several small
parameters characterizing the source of gravity. They are:
$\varepsilon_i\sim v_i/c$, $\varepsilon_e\sim v_e/c$, and $\eta_i\sim U_i/c^2$, $\eta_e\sim U_e/c^2$, where $v_i$ is
a characteristic velocity of motion of matter inside a body, $v_e$ is a characteristic velocity of the relative
motion of the bodies with respect to each other, $U_i$ is the internal gravitational potential of each body,
and $U_e$ is the external gravitational potential between the bodies. If one denotes a characteristic radius of
a body as $L$ and a characteristic distance between the bodies as $R$, the internal and external gravitational
potentials will be $U_i\simeq GM/L$ and $U_e\simeq GM/R$, where $M$ is a characteristic mass of the body.
Due to the virial theorem of the Newtonian gravity \citep{Chandr_1965ApJ142_1513} the small parameters are not independent.
Specifically, one has
$\varepsilon_i^2\sim\eta_i$ and $\varepsilon_e^2\sim\eta_e$. Hence,
parameters $\varepsilon_i$ and $\varepsilon_e$ are sufficient in doing post-Newtonian approximations. Because within
the solar system these parameters do not significantly differ from each other, they will be not distinguished
when doing the post-Newtonian iterations. In particular, notation $\varepsilon\equiv 1/c$ is used to mark the
powers of the fundamental speed $c$ in the post-Newtonian terms. This parameter is also considered as a primary
parameter of the PNA scheme to each all the other parameters are approximately equal,
for example, $\varepsilon_i=\varepsilon v_i$, $\eta_i=\varepsilon^2 U_i$, etc.

We work in the framework of general relativity and adopt the harmonic coordinates $x^\a=(x^0,x^i)$, where
$x^0=ct$, and $t$ is the coordinate time. The class of the harmonic coordinates is defined by imposing the de
Donder gauge condition on the metric tensor \citep{fock_1964book,weinberg_1972},
\be\la{har5}
\partial_\a\lt(\sqrt{-g}g^{\a\b}\rt)=0\;.
\ee
Our choice of the harmonic coordinates is not of the principal value. Calculations could be performed in arbitrary coordinates. Nonetheless the choice of the harmonic coordinates is dictated by their long-term use in relativistic celestial mechanics, astrometry and geodesy \citep{brumberg_1991_book,Kopeikin_2011_book,Kopeikin_2014_book1,Mueller_2008JGeod}. Furthermore, all relativistic algorithms of the data processing of high-precise astronomical and geodetic observations are written down in harmonic coordinates and are recommended to use by the International Astronomical Union (IAU) resolutions \citep{petit_2010,Soffel_2003AJ}

Einstein equations for the metric tensor are a complicated non-linear system of differential equations in
partial derivatives. Because gravitational field of the solar system is weak and motion of matter is slow, we
can focus on the first post-Newtonian approximation of general relativity. Furthermore, we assume that Earth
rotates uniformly with angular velocity $\o$ around a fixed axis which we identify with z-axis. We shall also
neglect tides, and consider Earth as an isolated body. Under these assumptions the spacetime becomes stationary
with the post-Newtonian metric having the following form \citep{Kopeikin_2011_book}
\ba\la{pnm1}
g_{00}&=&-1+\frac{2V}{c^2}+\frac2{c^4}\lt(\Phi-V^2\rt)+{\cal O}\lt({c^{-6}}\rt)\;,\\
g_{0i}&=&-\frac{4V^i}{c^3}+{\cal O}\lt(c^{-5}\rt)\;,\\
g_{ij}&=&\delta_{ij}\lt(1+\frac{2V}{c^2}\rt)+{\cal O}\lt(c^{-4}\rt)\;,
\ea
where the gravitational potentials entering the metric, satisfy the Poisson equations,
\ba\la{feq1}
\D V&=&-4\pi G\r\;,\\\la{feq2}
\D V^i&=&-4\pi G\r v^i\;,\\\la{feq3}
\D\Phi&=&-4\pi G\r\lt(2 v^2+2 V+\Pi+\frac{3p}{\e}\rt)\;,
\ea
with $p$ and $v^i$ being pressure and velocity of matter respectively, and $\Pi$ is the internal energy of
matter per unit mass. We emphasize that $\r$ is the local mass density of baryons per a unit of invariant
(3-dimensional) volume element $dV=\sqrt{-g}u^0d^3x$, where $u^0$ is the time component of the 4-velocity of
matter. The local mass density, $\rho$, relates in the post-Newtonian approximation to the invariant mass
density $\r^*=\sqrt{-g}u^0\rho$. In the first post-Newtonian approximation this equation reads
\be
\r^*=\r+\frac\r{c^2}\lt(\frac12 v^2+3V\rt)\;.
\ee
We assume that the matter consists of a perfect fluid. Then, the internal energy, $\Pi$, is related to
pressure, $p$, and the local density, $\rho$, by thermodynamic equation
\be\la{trm2}
d\Pi+pd\lt(\frac1\r\rt)=0\;,
\ee
and the equation of state, $p=p(\r)$. In the present paper we consider the case of a body consisting of a fluid
with a constant mass density $\rho={\rm const}$. Equation \eqref{trm2} states then, that inside such a fluid
the internal energy $\Pi$ is also constant.  

In the stationary spacetime, the mass density $\r^*$ obeys the {\it exact} equation of continuity
\be\la{vel3}
\partial_i\lt(\r^*v^i\rt)=0\;.
\ee
Velocity of rigidly rotating fluid is
\be\la{vel4}
v^i=\varepsilon^{ijk} \omega^j x^k\;,
\ee
where $\omega^i$ is the constant angular velocity. Replacing velocity $v^i$ in \eqref{vel3} with \eqref{vel4},
and differentiating, reveals that
\be\la{ily7}
v^i\partial_i \r=0\;,
\ee
which means that velocity of the fluid is tangent to the surfaces of constant density $\r$.

Modelling the real Earth as a rotating fluid ball of constant density is, of course, unrealistic from geophysical point of view as it is inconsistent neither with the seismological data \citep{Afanasiev_2016GeoJ} nor with IERS data on the Earth's rotation which clearly indicates the presence of the several layers of different density. Nonetheless, when one considers the geodetic problem of the terrestrial reference frame the realistic model of the Earth's interior leads to enormous practical difficulties in establishing the geoid's surface. Indeed, calculation of the geoid requires a well-defined reference level surface occupied by rotating fluid and approximating the geoid with the height's deviations as minimal as possible. If one takes a realistic mass distribution the reference level surface can not be an ellipsoid of rotation \citep{chandra_book}. Furthermore, the gravity field of such a figure of equilibrium (so-called, {\it normal gravity field } \citep{Moritz_1967}) will be described by too complicated mathematical equations which are not suitable for practical applications. A compromise is to take the reference level of the rotating fluid body as an ellipsoid which allows us to derive the {\it normal gravity field} in a concise meaningful form. However, such a reference ellipsoid of rotation can be maintained only by a homogeneous density distribution \citep{chandra_book}. Fortunately, the maximum deviation between the level surfaces of the realistic density distribution and the surfaces of equal density are of the order of $e^2\simeq 1/298$ only, and the differences in stress at the model remain considerably smaller than in the real Earth \citep{Torge_2012_book,Moritz_2006_book}. This explains why the classic geodesy operates with the reference Maclaurin ellipsoid of constant density as a reference surface. 

The same reasons are applied for justification of using the constant density $\r$ to build the reference level surface in relativistic geodesy. The constant density allows us to solve the post-Newtonian equations exactly so that we can write down precise mathematical equations to describe the gravitational field of the post-Newtonian rotating fluid configuration. However, as we shall see in the next section, the rotating fluid of constant density cannot be an ellipsoid of revolution in the post-Newtonian approximation but a surface of the fourth order. In the post-Newtonian approximation  it is possible to build the rotating ellipsoid of revolution which is a surface of the second order, only under assumption that the density of the fluid has an inhomogeneous ellipsoidal distribution of mass \citep{kop_2016}. The deviation from the inhomogeneity are of the post-Newtonian order of magnitude, $\d\r/\r\simeq GM_\oplus/c^2R_\oplus\simeq 7\times 10^{-10}$, and are practically unmeasurable in local experiments. Nonetheless, the relativistic effects in the normal gravity field produced by such a density inhomogeneity over the global scale might be noticed in precise measurements of the gravity field conducted with the next generation of gravimeters \citep{Imanishi_2004Sci,Flury_2004EMP} and/or gravity gradientometers \citep{Paik_1989AdSpR,Bagaev_2014RScI,Sugarbaker_2015APS}. Thus, we again has to compromise between two models of the Earth's interior - a non-homogeneous density distribution of matter inside a reference level ellipsoid or a homogeneous density with the small, post-Newtonian deviations of the reference level surface from the precise ellipsoid of revolution. It is not the goal of the present paper to decide between the two cases. Instead, we focus on the solution of the problem of the rotating fluid having a homogeneous distribution of density as the most tractable mathematical case extrapolating the Newtonian case of the Maclaurin ellipsoid to relativistic geodesy. The case of the inhomogeneous density distribution and comparison with the homogeneous density model will be considered somewhere else.

\section{Post-Newtonian reference-ellipsoid}\la{sec3}

In classical geodesy the reference figure for calculation of geoid's undulation is the Maclaurin ellipsoid of a
rigidly rotating fluid of a constant density $\rho$. Maclaurin's ellipsoid is a surface described by a polynomial of the second order
\citep{chandra_book}
\be\la{bv32}
\frac{x^2+y^2}{a^2}+\frac{z^2}{b^2}=1\;,
\ee
where $a$ and $b$ are semimajor and semiminor axes of the ellipsoid.
This property is because the differential Euler equation defining the
equilibrium of gravity and pressure is of the first order partial differential equation which first integral is the Newtonian gravity potential that is a scalar function represented by a polynomial of the second order with respect to the Cartesian spatial coordinates. In what follows, we assume $a>b$, and define the eccentricity of the Maclaurin ellipsoid by a standard formula \citep{Moritz_1967,Torge_2012_book}
\be\la{ecc4}
e\equiv\frac{\sqrt{a^2-b^2}}{a^2}\;. 
\ee

We shall demonstrate
in the following sections that in the post-Newtonian approximation the gravity potential, $W$, of the rotating
homogeneous fluid is a polynomial of the fourth order as was first noticed by Chandrasekhar \citep{Chandr_1965ApJ142_1513}. Hence, the level surface of a rigidly-rotating fluid is
expected to be a surface of the fourth order. We shall assume that the surface remains axisymmetric in the
post-Newtonian approximation and dubbed the body with such a surface as a PN ellipsoid \citep{footnote-1}.

We shall denote all quantities taken on the surface of the PN ellipsoid with a bar to distinguish them from the
coordinates outside of the surface.
Let the barred coordinates $\bar x^i=\{\bar x,\bar y,\bar z\}$ denote a point on the surface of the
PN ellipsoid with the axis of symmetry directed along the rotational axis and with the origin located at its
post-Newtonian center of mass. Post-Newtonian definitions of mass, the center of mass, and the other multipole
moments of an extended astronomical body can be found, for example, in \citep[chapter 4.5.3]{Kopeikin_2011_book} and are also given in section \ref{sec7} of the present paper. Let the rotational axis coincide with the direction of
$z$ axis. Then, the most general equation of the PN ellipsoid is
\be\la{1}
\frac{ \s^2}{a^2}+\frac{ z^2}{b^2}=1+\kappa F( { \bm x})\;,
\ee
where $\s^2\equiv  x^2+ y^2$, $\kappa\equiv \pi G\r a^2/c^2$ is the post-Newtonian parameter
which is convenient in the calculations that follow,
\be\la{1a}
F({\bm x})\equiv K_1\frac{ \s^2}{a^2}+K_2\frac{ z^2}{b^2}+ E_1\frac{ \s^4}{a^4}+E_2\frac{
z^4}{b^4}+E_3\frac{ \s^2  z^2}{a^2 b^2}\;,
\ee
and $K_1, K_2$, $E_i$ $(i=1,2,3)$ are arbitrary numerical coefficients. 

Let $x^i=\{x,y,z\}$ be any point inside the PN ellipsoid. We introduce a quadratic polynomial
\be\la{quadp98}
C({\bm x})\equiv\frac{\s^2}{a^2}+\frac{ z^2}{b^2}-1\;,
\ee 
where $\s^2\equiv x^2+y^2$.
This polynomial vanishes on the boundary surface of the Maclaurin ellipsoid \eqref{bv32}. However,
in the post-Newtonian approximation we have on the boundary of the PN ellipsoid \eqref{1} the following condition
\be\la{uui8}
C(\bar{\bm x})=\kappa F(\bar {\bm x})\;.
\ee
In terms of the polynomial $C( {\bm x})$ function $F( {\bm x})$ in the
right-hand side of \eqref{1} can be formally recast to
\be\la{1b}
F({\bm x})=K_1+E_1-\lt(K_1-K_2+2E_1-E_3\rt)\frac{ z^2}{b^2}+\lt(E_1+E_2-E_3\rt)\frac{ z^4}{b^4}+{\mathfrak
R}({\bm x})\;,
\ee
where the reminder
\be\la{rte1}
{\mathfrak R}( {\bm x})\equiv \lt[K_1+E_1+E_1\frac{ \s^2}{a^2}+\lt(E_3-E_1\rt)\frac{
z^2}{b^2}\rt]C({\bm x})\;.
\ee
The remainder ${\mathfrak R}( {\bm x})$ can be discarded on the boundary
of the PN ellipsoid because ${\mathfrak R}(\bar {\bm x})=0$. This property indicates a specific freedom in the definition of the surface of the  PN ellipsoid. Namely, equation \eqref{1} is defined up to a class of equivalence {\it modulo} function $C({\bm x})$ given in \eqref{quadp98} that vanishes on the surface of the Maclaurin ellipsoid, $C(\bar{\bm x})=0$. It means that the surface of the PN ellipsoid defined by equation \eqref{1} is always determined in the post-Newtonian terms only up to a function $\displaystyle\lt(\a_1\frac{ \s^2}{a^2}+\a_2\frac{z^2}{b^2}\rt)C({\bm x})$ where $\a_1,\a_2$ are arbitrary constant parameters. By making a specific choice of the parameters $\a_1$, $\a_2$ we can eliminate any two of the five coefficients $K_1,K_2,E_1,E_2,E_3$ entering function $F({\bm x})$ in \eqref{1}. It is convenient to chose $K_1=K_2=0$ that simplifies the equations which follow. The choice $K_1=K_2=0$ is equivalent to rescaling the semimajor and semiminor axes, $a$ and $b$, in \eqref{1} respectively.

Each cross-section of the PN ellipsoid being orthogonal to the rotational axis, represents a circle. The
equatorial cross-section has an equatorial radius, $\bar\s=r_e$, being determined from \eqref{1} by the
condition $\bar z=0$. It yields
\be\la{eqra}
r_e=a\lt(1+\frac12\kappa E_1\rt)\;.
\ee
The meridional cross-section of the PN ellipsoid is no longer an ellipse (as it was in case of the Maclaurin
ellipsoid) but a curve of the fourth order. Nonetheless, we can define the polar radius, $\bar z=r_p$, of the
PN ellipsoid by the condition, $\bar\s=0$. Equation \eqref{1} yields
\be\la{polra}
r_p=b\lt(1+\frac12\kappa E_2\rt)\;.
\ee
In terms of the parameters $r_e$ and $r_p$  equation \eqref{1} of the PN ellipsoid takes on the following form 
\be\la{mnh763}
\frac{\sigma^2}{r_e^2}+\frac{ z^2}{r_p^2}=1-\kappa\lt(E_1+E_2-E_3\rt)\lt(\frac{ z^2}{r_p^2}-\frac{ z^4}{r_p^4}\rt)\;.
\ee
This reveals that only a single combination, $E_1+E_2-E_3$, of the parameters explicitly appears in the description of the shape of the PN ellipsoid in the harmonic coordinates while the other two parameters, $E_1$ and $E_2$, can be absorbed (like the coefficients $K_1$ and $K_2$ above) to its equatorial and polar radii. The combination $E_1+E_2-E_3$ is determined by the physical equation of the equilibrium of the rotating fluid as explained in section \ref{sec8}. Parameters $E_1$ and $E_2$ are not limited by physics and can be chosen arbitrary within the accuracy allowed by the post-Newtonian approximation. We discuss their possible choice in sections \ref{sec8} and \ref{sec11}.

\begin{figure}
\begin{center}
\includegraphics[scale=0.3]{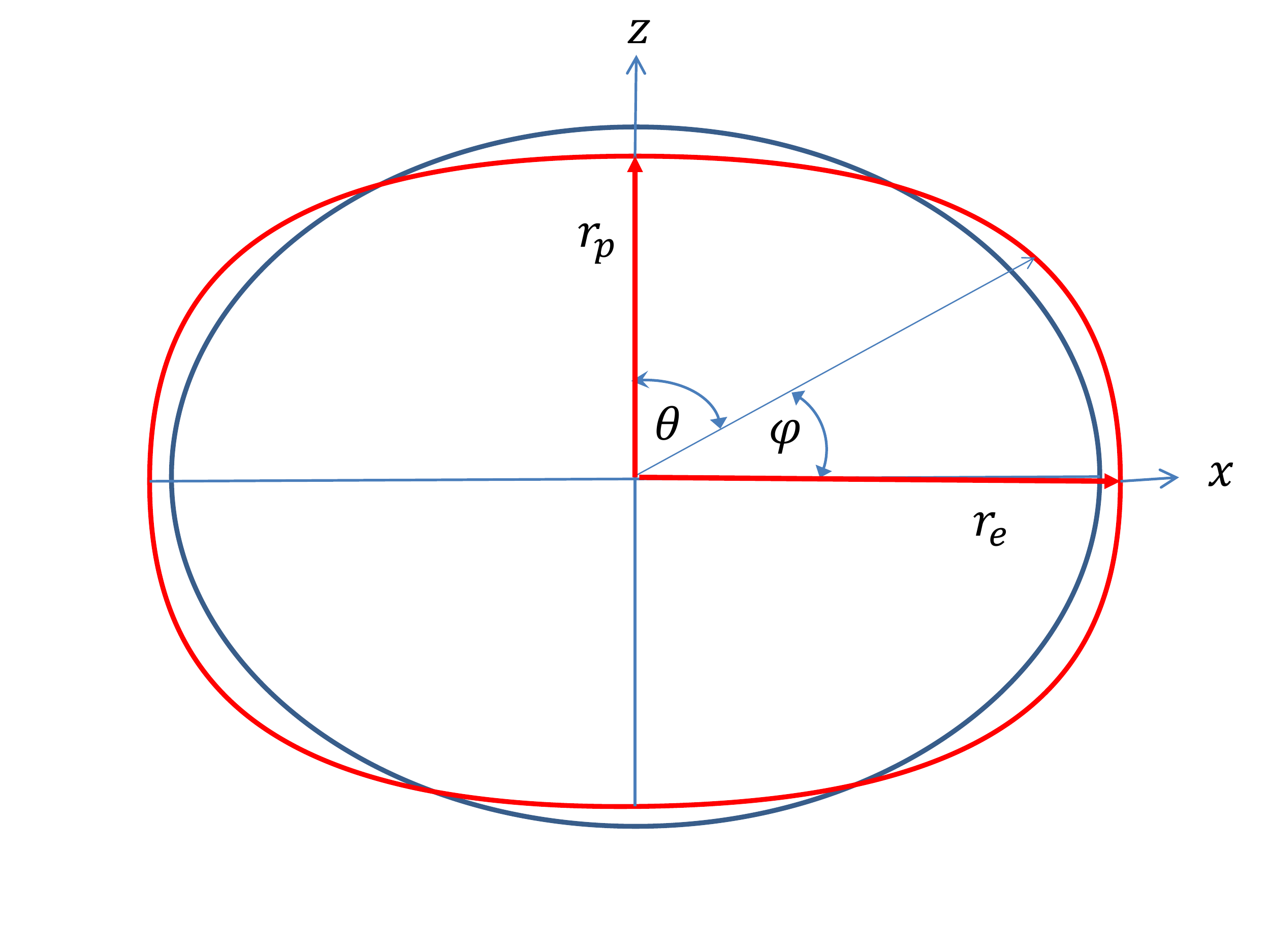}
\hfill
\includegraphics[scale=0.3]{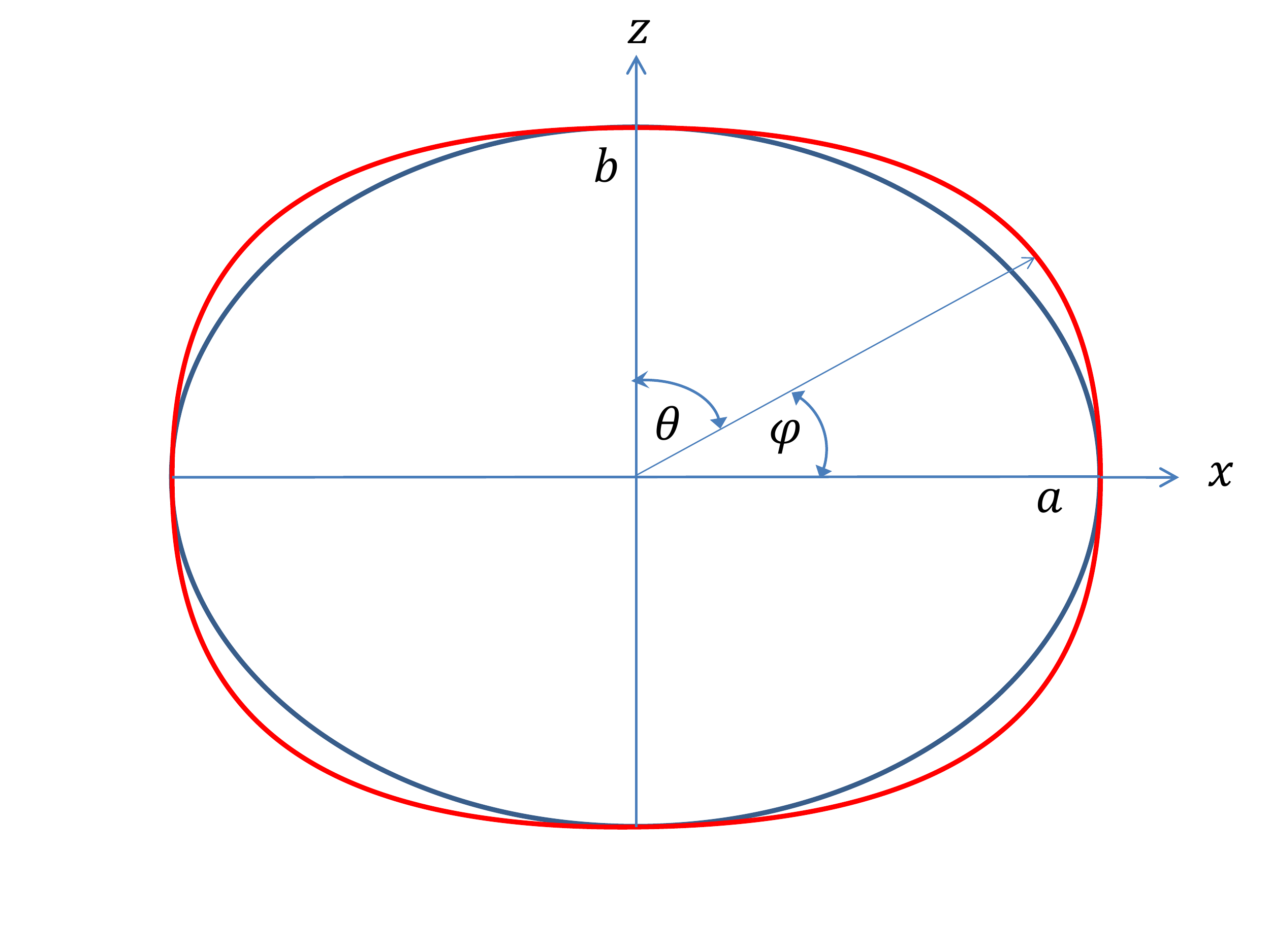}
\caption{Meridional cross-section of the PN ellipsoid (a red curve in the on-line version) versus the Maclaurin ellipsoid (a blue curve in the on-line version).
The left panel represents the most general case with arbitrary values of the shape parameters $E_1,E_2,E_3$
when the equatorial, $r_e$, and polar, $r_p$, radii of the PN ellipsoid differ from the semimajor, $a$, and
semiminor, $b$, axes of the Maclaurin ellipsoid, $r_e\not=a, r_p\not= b$. The right panel shows the case of $E_1=E_2=0$ when the equatorial and polar radii of the PN ellipsoid and the Maclaurin
ellipsoid are equal. The angle $\varphi$ is the geographic latitude ($-90^\circ\leq\varphi\leq 90^\circ$), and
the angle $\theta$ is a complementary angle (co-latitude) used for calculation of integrals in appendix of the
present paper ($0\leq\theta\leq\pi$). In general, when $E_1\not=E_2\not=0$, the maximal radial difference (the
'height' difference) between the surfaces of the PN ellipsoid and the Maclaurin ellipsoid depends on the choice
of the post-Newtonian coordinates, and can amount to a few cm. Carefully operating with the residual gauge
freedom of the post-Newtonian theory by choosing $E_1=E_2=0$, allows us to make the difference between the two
surfaces much less than one millimeter that is practically unobservable at the present time (for more discussion see section \ref{sec11}).
\label{fig1a}}
\end{center}
\end{figure}

We characterize the `oblateness' of the PN ellipsoid \eqref{mnh763} by the post-Newtonian `eccentricity'
\be\la{im76}
{\epsilon}\equiv\frac{\sqrt{r_e^2-r_p^2}}{r_e}\;.
\ee
It differs from the eccentricity \eqref{ecc4} of the Maclaurin ellipsoid by relativistic correction
\be\la{obt5}
{\epsilon}=e-\kappa\frac{1-e^2}{2e}\lt(E_2-E_1\rt)\;.
\ee
In case, when either $E_2=E_1$ or $E_1=E_2=0$, the two eccentricities coincide. 
The possible configurations of the PN ellipsoid versus Maclaurin's ellipsoid are
visualized in Fig. \ref{fig1a}.

\section{Post-Newtonian gauge freedom}\la{sec4}

Theoretical formalism for calculation of the post-Newtonian level surface can be worked out in arbitrary
coordinates. For mathematical and historic reasons the most convenient are harmonic coordinates which are also
used by the IAU \citep{Soffel_2003AJ} and IERS \citep{petit_2010} astro-geodetic data processing centers. The harmonic coordinates are selected by the
de Donder condition \eqref{har5} but it does not pick up a single coordinate system because of the property known as a residual gauge freedom \citep{Wald_1984_book}. It means that the gauge condition \eqref{har5} selects an infinite set of harmonic coordinates interrelating by coordinate transformations which don't violate the gauge condition \eqref{har5}. The field equations \eqref{feq1}--\eqref{feq3} and their solutions are form-invariant with respect to the residual gauge transformations. The residual gauge freedom can be further limited by imposing additional constraints on the metric tensor \citep{Kopeikin_2011_book}.

The residual gauge freedom of the harmonic coordinates is described by a post-Newtonian coordinate transformation,
\begin{equation}\la{tr56}
x'^\a=x^\a+\kappa\xi^\a(x)\;,
\end{equation}
where functions, $\xi^\a$, obey the Laplace equation \citep{Wald_1984_book},
\be\la{lap8}
\D\xi^\a=0\;.
\ee
We discuss the case of the solution of the post-Newtonian equations \eqref{feq1}--\eqref{feq3} inside the rotating fluid. Therefore, the solution of the Laplace equation \eqref{lap8} must be convergent at the origin of the coordinate system. It is well known that such a solution is given in terms
of the harmonic polynomials which are selected by the condition that the shape of the PN ellipsoid given by equation
\ref{1} remains the polynomial of the fourth order with the rotational symmetry about $z$-axis. This condition reduces the functions $\xi^\a$ in \eqref{tr56} to the
harmonic polynomials of the third order having the following form
\begin{subequations}
\ba\la{po98}
\xi^1&=&hx+\frac{px}{a^2}\lt(\s^2-4z^2\rt)\;,\\\la{po99}
\xi^2&=&hy+\frac{py}{a^2}\lt(\s^2-4z^2\rt)\;,\\\la{po23}
\xi^3&=&kz+\frac{qz}{b^2}\lt(3\s^2-2z^2\rt)\;,
\ea
\end{subequations}
where $h$, $k$, $p$ and $q$ are arbitrary constant parameters. It can be checked by direct inspection that the polynomials \eqref{po98}--\eqref{po23} satisfy the Laplace equation \eqref{lap8}. We have chosen, $\xi^0=0$, because we consider a stationary spacetime which means that
all functions are time-independent. We emphasize that the transformation \eqref{po98}--\eqref{po23} does not preserve the element of the coordinate volume $d^3x$ in the most general case. The coordinate volume would be preserved if $\pd_\a\xi^\a=0$ that implies $k=-2h$ and $q=-4b^2p/3a^2$. This constraint on the gauge transformation was originally employed by Chandrasekhar \citep{Chandr_1967ApJ147_334} who treated the volume-preserved transformations as the Lagrange displacements of the Maclaurin ellipsoid. Later on, Chandrasekhar abandoned this constraint \citep{Chandr_1971ApJ167_447} to adjust his theory to physical criteria for comparison of the post-Newtonian and Newtonian configurations of rotating fluids proposed by Bardeen \citep{Bardeen_1971ApJ} (see section \ref{sec8} for further detail).

Coordinate transformation \eqref{tr56} with $\xi^i$ taken from \eqref{po98}--\eqref{po23} does not violate the
harmonic gauge condition \eqref{har5} but it changes equations \eqref{1} and \eqref{1a} 
to 
\be\la{1kb4}
\frac{ \s'^2}{a^2}+\frac{ z'^2}{b^2}=1+\kappa F'( { \bm x}')\;,
\ee
\be\la{1am5d}
F'({\bm x}')\equiv K_1'\frac{ \s'^2}{a^2}+K_2'\frac{ z'^2}{b^2}+ E_1'\frac{ \s'^4}{a^4}+E_2'\frac{
z'^4}{b^4}+E_3'\frac{ \s'^2  z'^2}{a^2 b^2}\;,
\ee
where $\s'^2\equiv  x'^2+ y'^2$, and the primed coefficients are 
\begin{subequations}
\ba
\la{mk8b4}
K_1'&=&K_1+2h\;,\\
\la{mn34v} 
K_2'&=& K_2+2k\;,\\\la{zx5v} 
E_1'&=&E_1+2p\;,\\ 
E_2'&=&E_2-4q\;, \\
\la{bbby7} 
E_3'&=&E_3-8p\frac{b^2}{a^2}+6q\frac{a^2}{b^2}\;.\ea
\end{subequations}

Transformation equations \eqref{mk8b4}--\eqref{bbby7} make it evident that four out of the five coefficients $K_1$, $K_2$, $E_1,E_2,E_3$ are algebraically
independent. Moreover, there is one-to-one mapping between four parameters: $K_1 \leftrightarrow h$; $K_2\leftrightarrow k$; $E_1\leftrightarrow p$, and $E_2\leftrightarrow q$. It means that the choice of the coordinate parameters, $h$, $k$, $p$, and $q$, is actually equivalent to selecting the coefficients $K_1, K_2, E_1, E_2$ in the original equation \eqref{1} of the PN ellipsoid and fixing the residual gauge freedom of the harmonic coordinates. Because the geodetic data in classic geodesy is referred to the surface of the Maclaurin ellipsoid it would be practically useful to find such a post-Newtonian gauge in which the differences between the surfaces of the Maclaurin and PN ellipsoid were minimized. It would allow us to avoid unnecessary complications in adjusting the results of classic geodesy to the realm of general theory of relativity. Nonetheless, the question about what post-Newtonian gauge is the most convenient for geodesy remains open at the time being. We shall explore some possible options to fix the residual gauge in subsequent sections to see how the post-Newtonian physical equations defining the level surface, mass, angular momentum, etc., depend on the choice of the gauge in sections \ref{sec8} and \ref{sec11}.   

We have already fixed $K_1=K_2=0$. It complies with the transformations \eqref{mk8b4}, \eqref{mn34v} indicating that picking up the gauge parameters $h$, $k$ can always eliminate the coefficients $K_1$, $K_2$. We shall fix the coefficients $E_1,E_2$ later on, after solving the field equations and determining the gravitational potentials. The coefficient $E_3$ linearly depends on the choice of the parameters $p$ and $q$ and is truly gauge-dependent parameter. Its value is fixed (after choosing the parameters $E_1$ and $E_2$) by physics of the rotating fluid in the gravitational field leading to equation \eqref{eq1} of the equipotential level surface. 

Needless to say that physical quantities that make sense must be gauge-invariant quantities. In what follows we will demonstrate how to build the gauge invariant expressions for the total mass and angular momentum of the rotating fluid. Building the gauge-invariant expressions for the force of gravity is possible as well but takes us away from the canonical expressions adopted in the Newtonian geodesy. The gauge-invariant expressions for geodetic observables including the force of gravity are given, for example, in \citep{kop_2015PhLA,Oltean_2015arXiv}. However, the gauge-invariant approach in practical geodesy has little, if any application. Observables are gauge-invariant quantities but they are taken at different epochs and places, and must be interconnected. The interconnection of the observables is done with the help of the propagation equations mapping the observables to the fixed coordinate systems that are employed in fundamental astronomy and geodesy solely as the intermediate bookkeeper in order to compare the observables at one epoch to observables measured at another epoch. The process of the data processing maps one gauge-invariant quantity to another through the intermediate coordinate chart. This gauge-dependent chart is called reference ellipsoid, stellar fundamental catalogue, International Terrestrial Reference Frame (ITRF), etc., and they cannot be made gauge-independent because they are essentially realizations of the fundamental coordinate systems which are chosen and fixed by Conventions adopted at general assemblies of the IAU, IUGG, etc. Therefore, not all our results can be presented in the gauge-invariant form because this manuscript is about the fundamental coordinate systems in relativistic geodesy and their comparison.

\section{Newtonian Potential {\it V}}\la{sec5}

Newtonian gravitational potential $V$ satisfies the inhomogeneous Poisson's equation 
\be\la{poi1}
\D V({\bm x})=-4\pi G\r({\bm x})\;,
\ee
inside the mass. Its particular solution is given by 
\be\la{sol1}
V({\bm x})=\int_{\cal V}\frac{\r({\bm x}')d^3x'}{|{\bm x}-{\bm x}'|}\;,
\ee 
where ${\cal V}$ is the coordinate volume occupied by the matter distribution.
Inside the mass and under the assumption of the constant mass density $\r$, the integral \eqref{sol1} can be calculated by making
use of the spherical coordinates $\th,\l$ on a unit sphere. The procedure is as follows \citep{chandra_book}.

Let us consider a point $x^i=\{ x,y,z\}$ inside the PN ellipsoid \eqref{1}. It is connected to a point $\bar
x^i$ on the surface of the ellipsoid by a vector $R^i=\bar x^i-x^i$ where $R^i=R\ell^i$,
$R=\sqrt{\d_{ij}R^iR^j}$, and the unit vector, $\ell^i\equiv\{\sin\th\cos\l,\sin\th\sin\l,\cos\th\}$. In terms of these quantities we have
\be\la{3}
\bar x^i=x^i+\ell^iR\;.
\ee
Substituting \eqref{3} to \eqref{1} yields a quadratic equation
\be\la{4}
AR^2+2BR+C=\kappa F\lt({\bm x}+{\bm\ell}R\rt)\;,
\ee
where ${\bm x}\equiv \{x^i\}$, ${\bm l}=\{l^i\}$, and
\ba\la{5}
A\equiv\frac{\sin^2\th}{a^2}+\frac{\cos^2\th}{b^2}\;,\qquad
B\equiv \frac{\sin\th\lt(x\cos\l+y\sin\l\rt)}{a^2}+\frac{ z\cos\th}{b^2}\;,\qquad
C\equiv\frac{\s^2}{a^2}+\frac{ z^2}{b^2}-1\;.
\ea
We solve \eqref{4} iteratively by making use of $R=\hat R+c^{-2}\D R$, where $\hat R=(\hat R_+,\hat R_-)$
corresponds to two algebraically-independent solutions of the quadratic equation \eqref{4} with the right side being nil, and $\D R$ being yet unknown. After omitting terms of the order of ${\cal O}\lt(\kappa^2\rt)\simeq {\cal O}\lt(c^{-4}\rt)$, we have two roots 
\ba\la{6}
R_\pm&=&-\frac{B}{A}\pm\frac{\sqrt{B^2-AC+\kappa AF_\pm}}{A}\;,
\ea
where 
\ba\la{6a}
F_\pm&\equiv& E_1+\lt(E_3-2E_1\rt)\lt(\frac{ z+\cos\th\hat R_\pm}{b}\rt)^2+\lt(E_1+E_2-E_3\rt)\lt(\frac{
z+\cos\th\hat R_\pm}{b}\rt)^4+{\mathfrak R}\lt({\bm x}+{\bm\ell}R_\pm\rt)\;.
\ea

We make replacement of variable ${\bm x}'$ in \eqref{sol1} to ${\bm r}={\bm x}-{\bm x}'$, and use the spherical
coordinates to perform the integration with respect to the radial coordinate $r=|{\bm x}-{\bm x}'|$. After
integrating, the integral \eqref{sol1} takes on the following form \citep{chandra_book}
\be\la{7}
V=\frac14G\r\oint_{S^2}\lt(R_+^2+R_-^2\rt)d\Omega\;,
\ee
where $R_+$ and $R_-$ are defined in \eqref{6}.
After making use of \eqref{6} and expanding the integrand in \eqref{7} with respect to the post-Newtonian
parameter $\kappa$, the Newtonian potential takes on the following form
\be\la{8}
V=\frac12G\r\oint_{S^2}\lt\{\frac{2B^2-AC}{A^2}+\frac{\kappa}{2A}\lt[F_{+}+F_{-}-\frac{B}{\sqrt{B^2-AC}}\lt(F_{+}-F_{-}\rt)\rt]\rt\}d\Omega\;,
\ee
where all post-Newtonian terms of the higher order with respect to $\kappa$ have been discarded, the integration is
performed over a unit sphere $S^2$ with respect to the angles $\lambda$ and $\theta$, and $d\Omega\equiv\sin\theta
d\theta d\lambda$ is the element of the solid angle on the unit sphere.

Now, we expand $F_\pm$ in a polynomial w.r.t. $R_\pm$,
\be\la{9}
F_\pm=\a_0+\a_1 \cos\th R_\pm+\a_2 \lt(\cos\th R_\pm\rt)^2+\a_3\lt(\cos\th R_\pm\rt)^3+\a_4 \lt(\cos\th
R_\pm\rt)^4\;,
\ee
where the residual term ${\mathfrak R}_\pm$ vanishes because it is proportional to $C(\bar{\bm
x})=AR^2+2BR+C=0+{\cal O}(\kappa)$, and the coefficients
\ba\la{10}
\a_0&=&E_1+\lt(E_3-2E_1\rt)\frac{ z^2}{b^2}+\lt(E_1+E_2-E_3\rt)\frac{ z^4}{b^4} \;,   \\\la{10a}
\a_1&=&\frac{2 z}{b^2}\lt[E_3-2E_1+2\lt(E_1+E_2-E_3\rt)\frac{ z^2}{b^2}\rt]      \\\la{10b}
\a_2&=& \frac{1}{b^2}\lt[E_3-2E_1+6\lt(E_1+E_2-E_3\rt)\frac{ z^2}{b^2}\rt]     \\\la{10c}
\a_3&=& \frac{4 z}{b^4}\lt(E_1+E_2-E_3\rt)\;,     \\\la{10d}
\a_4&=& \frac{1}{b^4}\lt(E_1+E_2-E_3\rt)\;,  
\ea
are polynomials of $z$ only. We also notice that on the surface of the PN ellipsoid, $F(\bar{ \bm
x})=\a_0$, as follows from \eqref{1b} and \eqref{rte1}. We can also use, $C(\bar{\bm x})=0$, in the
post-Newtonian terms.

Replacing \eqref{9} in \eqref{8} transforms it to
\be\la{gg1}
V=V_N+\kappa V_{pN}\;,
\ee
where
\ba\la{11aa}
V_N&=&G\r\oint_{S^2}\lt(\frac{B^2}{A^2}-\frac{C}{2A}\rt)d\Omega\\\la{11bb}
V_{pN}&=&G\r\oint_{S^2}
\lt[\frac{\a_0}{2A}-\a_1\cos\th\frac{B}{A^2}+\a_2\cos^2\th\lt(\frac{2B^2}{A^3}-\frac{C}{2A^2}\rt)\rt.\\\nonumber&&\quad\qquad-2\a_3\cos^3\th\lt(\frac{2B^3}{A^4}-\frac{BC}{A^3}\rt)+\lt.\a_4\cos^4\th\lt(\frac{8B^4}{A^5}-\frac{6B^2C}{A^4}+\frac{C^2}{2A^3}\rt)\rt]d\Omega\;.
\ea
Equations \eqref{gg1}--\eqref{11bb} describe the Newtonian potential {\it exactly} both on the surface of
the PN ellipsoid and inside it.

The integrals in \eqref{11aa}, \eqref{11bb} are discussed in Appendix \ref{appen1}. After
evaluating the integrals and reducing similar terms,
potentials $V_N$ and $V_{pN}$ take on the following form:
\ba\la{nn1}
V_N&=&\pi G\r a^2\lt[\lt(1 - \frac{z^2}{b^2}\rt){\gimel}_0- \lt(1 - 3\frac{z^2}{b^2}\rt){\gimel}_1-C({\bm
x}){\gimel}_1\rt]\;,\\\la{nn2}
V_{pN}&=&\pi G\r a^2\lt[F_1(z)+b^2F_2(z)C({\bm x})+b^4F_3(z)C^2({\bm x})\rt]\;,
\ea
where
\ba\la{qq1}
F_1(z)&=&\a_0 {\gimel}_0-2\a_1 z {\gimel}_1+\\\nonumber  
&&2\a_2b^2\lt[\lt(1-\frac{z^2}{b^2}\rt){\gimel}_1-\lt(1-\frac{3z^2}{b^2}\rt){\gimel}_2\rt]-4\a_3 b^2
z\lt[3\lt(1-\frac{z^2}{b^2}\rt){\gimel}_2-\lt(3-\frac{5z^2}{b^2}\rt){\gimel}_3\rt]+\\\nonumber
&&6\a_4b^4\lt[
\lt(1-\frac{z^2}{b^2}\rt)^2{\gimel}_2-2\lt(1-6\frac{z^2}{b^2}+5\frac{z^4}{b^4}\rt){\gimel}_3+\lt(1-10\frac{z^2}{b^2}+\frac{35}{3}\frac{z^4}{b^4}\rt){\gimel}_4
\rt]\;,
\\\la{qq2}
F_2(z)&=&\a_2\lt({\gimel}_1-2 {\gimel}_2\rt)-4\a_3 z\lt(2{\gimel}_2-3 {\gimel}_3\rt)+\\\nonumber
&&6\a_4
b^2\lt[\lt(1-\frac{z^2}{b^2}\rt){\gimel}_2-3\lt(1-\frac{3z^2}{b^2}\rt){\gimel}_3+2\lt(1-\frac{5z^2}{b^2}\rt){\gimel}_4\rt]\;,\\\la{qq3}
F_3(z)&=&\a_4\lt({\gimel}_2-6{\gimel}_3+6{\gimel}_4\rt)\;,
\ea
and the polynomial coefficients $\a_0,\a_1,\a_2,\a_3,\a_4$ are given in \eqref{10}-\eqref{10d}. It is worth
noticing that the potential $V_N$ satisfies the Poisson equation \eqref{poi1} {\it exactly}. It means that the
post-Newtonian function $V_{pN}$ obeys the Laplace equation
\be\la{lap1}
\D V_{pN}=0\;,
\ee
and the right side of equation \eqref{nn2} is a harmonic polynomial of the fourth order.

\section{Post-Newtonian Potentials}\la{sec6}
\subsection{Vector Potential $V^i$}\la{sec6.1}
Vector potential $V^i$ obeys the Poisson equation
\be\la{pom7}
\D V^i=-4\pi G\r({\bm x}) v^i({\bm x})\;,
\ee
which has a particular solution
\be\la{potui}
V^i=G\int_{\cal V}\frac{\r({\bm x}') v^i(x')}{|{\bm x}-{\bm x}'|}d^3x'\;.
\ee
For a rigidly rotating configuration, $v^i(x)=\varepsilon^{ijk}\omega^j x^k$ so that
\be\la{kui8}
V^i=\varepsilon^{ijk}\omega^j{\cal D}^k\;,
\ee
where 
\be\la{om7}
{\cal D}^i=G\int\frac{\r({\bm x}') x'^i d^3x'}{|{\bm x}-{\bm x}'|}\;.
\ee 
It can be recast to the following form
\be\la{p9r}
{\cal D}^i=x^i V_N+G\int\frac{\r({\bm x}') (x'^i-x^i)}{|{\bm x}-{\bm x}'|}d^3x'\;,
\ee
where $V_N$ is the Newtonian potential given in \eqref{nn1}. For the case of a constant density, $\r({\bm
x}')=\r={\rm const}.$, the second term in the right hand side of \eqref{p9r} can be integrated over the radial
coordinate, yielding
\ba\la{wdv6}
\int_{\cal V}\frac{\r({\bm x}') (x'^i-x^i)}{|{\bm x}-{\bm
x}'|}d^3x'&=&\frac{\r}6\oint_{S^2}\lt(R^3_{+}+R^3_{-}\rt)l^id\Omega\;.
\ea
After making use of \eqref{6} to replace $R_+$ and $R_-$, we obtain
\ba\la{m95}
\int_{\cal V}\frac{\r (x'^i-x^i)}{|{\bm x}-{\bm x}'|}d^3x'
&=&\r\oint_{S^2}\lt(-\frac43\frac{B^3}{A^3}+\frac{BC}{A^2}\rt)l^id\Omega\;,
\ea
where we have omitted the post-Newtonian terms being proportional to ${\cal O}(\kappa)$ since the vector potential $V^i$ itself appears
only in the post-Newtonian terms.
Integrals entering \eqref{m95} are given in Appendix \ref{appen1}.
Calculation reveals
\ba
\int_{\cal V}\frac{\r({\bm x}') (x'-x)}{|{\bm x}-{\bm x}'|}d^3x'&=&-\pi \r a^2 x
\lt[\lt(1-\frac{z^2}{b^2}\rt){\gimel}_0-2\lt(1-\frac{3z^2}{b^2}\rt){\gimel}_1+\lt(1-\frac{5z^2}{b^2}\rt){\gimel}_2
\rt]+x\pi \r a^2 C({\bm x})\lt({\gimel}_1-{\gimel}_2\rt)\;,\\
\int_{\cal V}\frac{\r ({\bm x}')(y'-y)}{|{\bm x}-{\bm x}'|}d^3x'&=&-\pi \r a^2 y
\lt[\lt(1-\frac{z^2}{b^2}\rt){\gimel}_0-2\lt(1-\frac{3z^2}{b^2}\rt){\gimel}_1+\lt(1-\frac{5z^2}{b^2}\rt){\gimel}_2
\rt]+y\pi \r a^2 C({\bm x})\lt({\gimel}_1-{\gimel}_2\rt)\;,\\
\int_{\cal V}\frac{\r({\bm x}') (z'-z)}{|{\bm x}-{\bm x}'|}d^3x'&=&-4\pi \r a^2z
\lt[\lt(1-\frac{z^2}{b^2}\rt){\gimel}_1-\lt(1-\frac{5z^2}{3b^2}\rt){\gimel}_2\rt]-2z\pi \r a^2 C({\bm
x})\lt({\gimel}_1-2{\gimel}_2\rt)\;.
\ea
Substituting this result to \eqref{p9r} and making use of \eqref{nn1} yields 
\be\la{kui9}
{\cal D}^i\equiv\lt({\cal D}^x,{\cal D}^y,{\cal D}^z\rt)=\lt(xD_1,yD_1,zD_2\rt)\;,
\ee
where functions
\ba\la{u76}
D_1&\equiv&\pi G\r a^2\lt[\lt(1-\frac{3z^2}{b^2}\rt){\gimel}_1-\lt(1-\frac{5z^2}{b^2}\rt){\gimel}_2-C({\bm
x}){\gimel}_2\rt]\;,\\\la{u77}
D_2&\equiv&\pi G\r
a^2\lt[\lt(1-\frac{z^2}{b^2}\rt){\gimel}_0-\lt(5-7\frac{z^2}{b^2}\rt){\gimel}_1+4\lt(1-\frac{5z^2}{3b^2}\rt){\gimel}_2+C({\bm
x})\lt(4{\gimel}_2-3{\gimel}_1\rt)\rt]\;.
\ea
\subsection{Scalar Potential $\Phi$}\la{sec6.2}

Potential $\Phi$ is defined by equation
\be\la{eqphi}
\D\Phi=-4\pi G\r({\bm x}')\phi({\bm x}')\;,
\ee
where function
\be
\phi({\bm x}')\equiv 2\omega^2\s^2+3\frac{p}{\r}+2V_N\;.
\ee
In the Newtonian approximation pressure $p$ inside the massive body with a constant density $\r$ has an ellipsoidal distribution and is given by solution of the equation of a hydrostatic equilibrium,
\citep{chandra_book}
\be\la{c65}
\frac{p}{\r}=-\pi G\r a^2C({\bm x})\lt(\gimel_0-2\gimel_1\rt)\;.
\ee
Making use of \eqref{c65} 
and \eqref{nn1} 
we can write down function $\phi({\bm x}')$ as
\be
\phi({\bm x}')=a^2\lt[2\omega^2-\pi
G\r\lt(3\gimel_0-4\gimel_1\rt)\rt]\left(\frac{\s^2}{a^2}+\frac{z^2}{b^2}\rt)-
2a^2\lt[\omega^2-\pi G\r\lt(\gimel_0-3\gimel_1\rt)\rt]\frac{z^2}{b^2}+\pi G\r a^2\lt(5\gimel_0-6\gimel_1\rt)\;.
\end{equation}
Particular solution of \eqref{eqphi} can be written, then, as
\be\la{ph23}
\Phi=2\omega^2 a^2\lt(I_1-I_2\rt)-\pi G\r
a^2\lt[\lt(3I_1+2I_2-5V_N\rt)\gimel_0+2\lt(2I_1+3I_2-3V_N\rt)\gimel_1\rt]
\;,
\ee
where we have introduced two new integrals
\be
I_1=G\r\int_{\cal V}\frac{ d^3x'}{|{\bm x}-{\bm x}'|}\lt(\frac{\s'^2}{a^2}+\frac{z'^2}{b^2}\rt) \;,\qquad\qquad
I_2=\frac{G\r}{b^2}\int_{\cal V}\frac{ z'^2}{|{\bm x}-{\bm x}'|} d^3x'\;.
\ee
The integrals can be split in several algebraic pieces,
\ba\la{i1vv}
I_1&=&\lt(\frac{\s^2}{a^2}+\frac{z^2}{b^2}\rt)\lt(2D_1-V_N\rt)+2\frac{z^2}{b^2}\lt(D_2-D_1\rt)
+\frac{G\r}{8}\oint_{S^2}\lt(R^4_++R^4_-\rt)\lt(\frac{\sin^2\th}{a^2}+\frac{\cos^2\th}{b^2}\rt)
d\Omega\;,\\\la{i1nn}
I_2&=&\frac{z^2}{b^2}\lt(2D_2-V_N\rt)+\frac{G\r}{8b^2}\oint_{S^2}\lt(R^4_++R^4_-\rt)\cos^2\th d\Omega\;,
\ea
where the integrals
\ba\la{i1as}
\frac{1}{8}\oint_{S^2}\lt(R^4_++R^4_-\rt)\lt(\frac{\sin^2\th}{a^2}+\frac{\cos^2\th}{b^2}\rt)
d\Omega&=&\oint_{S^2}\lt(\frac{2B^4}{A^3}-2\frac{B^2 C}{A^2}+\frac{C^2}{4A}\rt) d\Omega\;,\\\la{i1an}
\frac{1}{8}\oint_{S^2}\lt(R^4_++R^4_-\rt)\cos^2\th d\Omega&=&\oint_{S^2}\lt(\frac{2B^4}{A^4}-2\frac{B^2
C}{A^3}+\frac{C^2}{4A^2}\rt)\cos^2\th d\Omega\;.
\ea
We use the results of Appendix \ref{appen1} to calculate these integrals, and obtain
\ba \la{i1}
\oint_{S^2}\lt(\frac{2B^4}{A^3}-2\frac{B^2 C}{A^2}+\frac{C^2}{4A}\rt) d\Omega
&=&\frac32\pi
a^2\lt[\lt(1-\frac{z^2}{b^2}\rt)^2{\gimel}_0-2\lt(1-6\frac{z^2}{b^2}+5\frac{z^4}{b^4}\rt){\gimel}_1+\lt(1-10\frac{z^2}{b^2}+\frac{35}{3}\frac{z^4}{b^4}\rt){\gimel}_2\rt]\\\nonumber
&&+\pi
a^2\lt[\lt(1-\frac{z^2}{b^2}\rt){\gimel}_0-4\lt(1-3\frac{z^2}{b^2}\rt){\gimel}_1+3\lt(1-5\frac{z^2}{b^2}\rt){\gimel}_2\rt]C({\bm
x})\\\nonumber
&&-\pi  a^2\lt[{\gimel}_1-\frac32 {\gimel}_2 \rt]C^2({\bm x})\;,
\ea
\ba\la{i2}
\oint_{S^2}\lt(\frac{2B^4}{A^4}-2\frac{B^2 C}{A^3}+\frac{C^2}{4A^2}\rt)\frac{\cos^2\th}{b^2} d\Omega
&=&\frac32\pi
a^2\lt[\lt(1-\frac{z^2}{b^2}\rt)^2{\gimel}_1-2\lt(1-6\frac{z^2}{b^2}+5\frac{z^4}{b^4}\rt){\gimel}_2+\lt(1-10\frac{z^2}{b^2}+\frac{35}{3}\frac{z^4}{b^4}\rt){\gimel}_3\rt]\\\nonumber
&&+\pi
a^2\lt[\lt(1-\frac{z^2}{b^2}\rt){\gimel}_1-4\lt(1-3\frac{z^2}{b^2}\rt){\gimel}_2+3\lt(1-5\frac{z^2}{b^2}\rt){\gimel}_3\rt]C({\bm
x})\\
&&+\pi  a^2\lt[{\gimel}_2-\frac32 {\gimel}_3 \rt]C^2({\bm x})\;.\nonumber
\ea
Substituting these results in \eqref{i1vv} and \eqref{i1nn} yields
\ba\la{pon5}
I_1&=&\frac12\pi G\r
a^2\lt[\lt(1-\frac{z^4}{b^4}\rt)\gimel_0-6\frac{z^2}{b^2}\lt(1-\frac{5}3\frac{z^2}{b^2}\rt)\gimel_1-\lt(1-10\frac{z^2}{b^2}+\frac{35}3\frac{z^4}{b^4}\rt)\gimel_2\rt]\\\nonumber
&&-\pi G\r a^2\lt[\frac{3z^2}{b^2}\gimel_1+\lt(1-5\frac{z^2}{b^2}\rt)\gimel_2+\frac12\gimel_2 C({\bm
x})\rt]C(\bm x)\;,\\\la{pon7}
I_2&=&\pi G\r
a^2\lt[\lt(1-\frac{z^2}{b^2}\rt)\frac{z^2}{b^2}\gimel_0+\lt(\frac32-12\frac{z^2}{b^2}+\frac{25}2\frac{z^4}{b^4}\rt)\gimel_1-\lt(3-26\frac{z^2}{b^2}+\frac{85}{3}\frac{z^4}{b^4}\rt)\gimel_2+\lt(\frac32-15\frac{z^2}{b^2}+\frac{35}2\frac{z^4}{b^4}\rt)\gimel_3\rt]\\\nonumber
&&+\pi G\r
a^2\lt[\lt(1-6\frac{z^2}{b^2}\rt)\gimel_1-4\lt(1-5\frac{z^2}{b^2}\rt)\gimel_2+3\lt(1-5\frac{z^2}{b^2}\rt)\gimel_3+\lt(\gimel_2-\frac32\gimel_3\rt)
C({\bm x})\rt]C({\bm x})\;.
\ea
Replacing these expressions to \eqref{ph23} results in
\be\la{phi23a}
\Phi=\Phi_0+\Phi_1 C({\bm x})+\Phi_2 C^2({\bm x})\;,
\ee
where
\ba\la{phi23b}
\Phi_0&=&\frac{1}{2}\pi^2 G^2\r^2 a^4 \lt[7 \gimel_0^2 - 3 \gimel_0 (8 \gimel_1 - 5 \gimel_2 + 2 \gimel_3) + 2
\gimel_1 (15 \gimel_1 - 20 \gimel_2 + 9 \gimel_3)\rt] \\\nonumber
& &-\pi^2 G^2\r^2 a^4\lt[7 \gimel_0^2 + \gimel_0 (-60 \gimel_1 + 67 \gimel_2 - 30 \gimel_3) + 2 \gimel_1 (51
\gimel_1 - 88 \gimel_2 + 45 \gimel_3)\rt]\frac{z^2}{b^2} \\\nonumber
&&+\frac{1}{6}\pi^2 G^2\r^2 a^4\lt[21 \gimel_0^2 + \gimel_0 (-288 \gimel_1 + 445 \gimel_2 - 210 \gimel_3) + 10
\gimel_1 (57 \gimel_1 - 116 \gimel_2 + 63 \gimel_3)\rt] \frac{z^4}{b^4}\\\nonumber
&&+\pi G\r a^4\omega^2\lt[\gimel_0 - 3 \gimel_1 + 5 \gimel_2 - 3 \gimel_3 - 
 2 \lt(\gimel_0 - 9 \gimel_1 + 21 \gimel_2 - 15 \gimel_3\rt)\frac{ z^2}{b^2} + \lt(\gimel_0 - 
    5 (3 \gimel_1 - 9 \gimel_2 + 7 \gimel_3)\rt) \frac{z^4}{b^4}\rt]\;,\\
\la{phi23c}
\Phi_1&=&\pi^2 G^2\r^2 a^4 \lt[\gimel_0 (-7 \gimel_1 + 11 \gimel_2 - 6 \gimel_3) + 
         2 \gimel_1 (6 \gimel_1 - 14 \gimel_2 + 9 \gimel_3)\rt] \\\nonumber
         &&+\pi^2 G^2\r^2 a^4 \lt[\gimel_0 (21 \gimel_1 - 55 \gimel_2 + 
            30 \gimel_3) - 2 \gimel_1 (24 \gimel_1 - 70 \gimel_2 + 45 \gimel_3)\rt]\frac{ z^2}{b^2}\\\nonumber
             &&-2\pi G\r a^4\omega^2 
   \lt[(\gimel_1 - 3 \gimel_2 + 3 \gimel_3) - 3 (\gimel_1 - 5 \gimel_2 + 5 \gimel_3)\frac{ z^2}{b^2}\rt]\;,\\
\la{phi23d}
\Phi_2&=&\frac12\pi G\r a^4\lt[(-\gimel_0 \gimel_2 + 8 \gimel_1 \gimel_2 + 6 \gimel_0 \gimel_3 - 
      18 \gimel_1 \gimel_3) \pi G\r -6 (\gimel_2 - \gimel_3) \omega^2\rt]\;.
\ea
This finalizes the calculation of the post-Newtonian potentials inside the rotating fluid body.

\section{Conserved Quantities}\la{sec7}
The post-Newtonian conservation laws have been discussed by a number of researchers, the most notably in
textbooks \citep{fock_1964book,will_1993book,Kopeikin_2011_book}. General relativity predicts that the
integrals of energy, linear momentum, angular momentum and the center of mass of an isolated system are
conserved in the post-Newtonian approximation. In the present paper we are dealing with a single isolated body so that
the integrals of the center of mass and the linear momentum are trivial, and we can always chose the origin of
the coordinate system at the center of mass of the body with the linear momentum being nil. The integrals of
energy and angular momentum are less trivial and requires detailed calculations which are given below.
\subsection{Post-Newtonian Mass}\la{sec7a}
The law of conservation of energy yields the post-Newtonian mass of a rotating fluid ball that is defined as
follows \citep{fock_1964book,will_1993book,Kopeikin_2011_book}
\be\la{totm1}
M=M_N+\frac1{c^2}M_{pN}\;,
\ee
where  
\be\la{newmas}
M_N=\int_{\cal V}\r({\bm x}) d^3x \;,
\ee
is the Newtonian mass of baryons comprising the body, 
\be\la{pnm123}
M_{pN}=\int_{\cal V}\r({\bm x}) \lt( v^2+\Pi+\frac52 V_N\rt)d^3x \;,
\ee
is the post-Newtonian correction taking into account the contribution of the internal kinetic, gravitational and compressional energies, and ${\cal V}$ is the coordinate volume of the PN ellipsoid. In what follows, we shall formally include the compressional energy $\Pi$ to the density $\r$ because $\Pi$ is constant.

Under condition that the density $\r({\bm x})=\r={\rm const}$, the rest mass is reduced to
\be
M_N=\r\,{\cal V}\;.
\ee
In order to calculate the volume, ${\cal V}$, we introduce the normalized spherical coordinates $r,\th,\lambda$
related to the Cartesian (harmonic) coordinates $x,y,z$ as follows,
\be\la{newc3}
x=ar\sin\th\cos\lambda\;,\qquad y=ar\sin\th\sin\lambda\;,\qquad z=br\cos\th\;.
\ee 
In these coordinates the volume ${\cal V}$ is given by
\be\la{ma1}
{\cal V}=a^2b\int\limits_0^{r(\th)}\int\limits_0^\pi\int\limits_0^{2\pi}r^2\sin\th dr d\th d\lambda\;,
\ee
where $r(\th)$ describes the surface of the PN ellipsoid defined above in \eqref{1} 
\be\la{ppn8}
r^2(\th)=1+\kappa\lt(E_1\sin^4\th+E_2\cos^4\th+E_3\sin^2\cos^2\th\rt)\;.
\ee
Integration in \eqref{ma1} results in 
\be\la{lov6}
M_N=\frac{4\pi}{3} \r a^2b\lt[1+\frac{\kappa}{10}\lt(8E_1+3E_2+2E_3\rt)\rt]\;,
\ee
which clearly indicates that the Newtonian mass, $M_N$, depends on the particular choice of the shape of the PN ellipsoid through the linear combination of the coefficients $E_1,E_2,E_3$.  

The post-Newtonian contribution, $M_{pN}$, to the rest mass reads
\ba\la{pmbx5}
M_{pN}&=&\r a^4b\int\limits_0^1\int\limits_0^\pi\int\limits_0^{2\pi}\lt\{\omega^2r^2\sin^2\th+\frac52\pi G\r
\lt[{\gimel}_0\lt(1-r^2\cos^2\th\rt)-{\gimel}_1r^2\lt(1-3\cos^2\th\rt)\rt]\rt\}r^2\sin\th dr d\th
d\lambda\\\nonumber
&=&\frac{8\pi}{15} \r a^4b\lt(\omega^2+5\pi G\r{\gimel}_0\rt)\;,
\ea
which was obtained from \eqref{pnm123} upon substitution of $v^2=\o^2r^2\sin^2\th$, and $V_N$ from \eqref{nn1}.
After adding up formulas \eqref{lov6} and \eqref{pmbx5} the total mass \eqref{totm1} becomes
\be\la{totm0}
M=\frac{4\pi}{3} \r a^2b\lt[1+\frac{\kappa}{10}\lt(\frac{4\o^2}{\pi G\r}+8E_1+3E_2+2E_3+20{\gimel}_0\rt)\rt]\;.\ee

Some clarifications are required at this point to prevent confusion with the residual gauge freedom described by equations \eqref{mk8b4}--\eqref{bbby7}, and the constancy of the total mass $M$ as the integral of motion of the fluid. If we do calculations in the primed harmonic coordinates, $x'^\a$, related to the original coordinates $x^\a$ by equation \eqref{tr56}, it changes the mathematical expression for the Newtonian mass
\be\la{newgt}
M_N=\int_{\cal V'}\r'({\bm x}')J(x') d^3x' \;,
\ee
where ${\cal V'}$ is the coordinate volume occupied by the same amount of mass in the primed coordinates, $\r'({\bm x}')=\r({\bm x})=\r$ is the constant mass density, $d^3x'$ is an element of the coordinate volume in the primed coordinates and
\be
J(x')={\rm det}\left|\frac{\pd x^i}{\pd x'^j}\right|=1-\pd_i\xi^i=1-\kappa\lt(\frac{4p}{a^2} + \frac{3q}{b^2}\rt)\left(x'^2 + y'^2 - 2 z'^2\right)\;,
\ee
is the Jacobian of the inverse coordinate transformation \eqref{tr56} with $h=k=0$. Integration in \eqref{newgt} with the volume bounded by the surface \eqref{1kb4} of the PN ellipsoid in the primed coordinates (that is the same equation \eqref{1kb4} but with the radial coordinate $r'=(x'^2+y'^2+z'^2)^{1/2}$ and the coefficients $E'_1,E'_2,E'_3$) yields the gauge-invariant expression for the Newtonian mass of the rotating fluid body
\be
M_N=\frac{4\pi}{3} \r a^2b\lt[1+\frac{\kappa}{10}\lt(8E'_1+3E'_2+2E'_3-16p+12q+16p\frac{b^2}{a^2}-12q\frac{a^2}{b^2}\rt)\rt]\;.
\ee
This expression naturally coincides with that given in \eqref{lov6} after making use of equations \eqref{zx5v}-\eqref{bbby7}. The post-Newtonian contribution, $M_{pN}$, to the total mass is not sensitive to the post-Newtonian coordinate transformation, and remains the same as in \eqref{pmbx5}. Therefore, the total mass is the gauge-invariant quantity under condition that the coefficients $E_1$ and $E_2$ have been fixed in a particular coordinate system. It is also worth mentioning that the combination of the residual gauge parameters, $-16p+12q+16pb^2/a^2-12qa^2/b^2=0$, when the residual gauge transformation \eqref{tr56} preserves the coordinate volume of integration, that is, when $\pd_i\xi^i=0$. 

Parameters $E_1$ and $E_2$ define the shape of the PN ellipsoid as compared with the shape of the Maclaurin ellipsoid in the chosen coordinate system. Picking up the shape of the PN ellipsoid is equivalent to eliminating the residual gauge freedom. Depending on their choice we have different options, for example, we can either equate the relativistic mass $M$ of the PN ellipsoid to the Newtonian mass of the Maclaurin ellipsoid (Bardeen-Chandrasekhar's gauge discussed at the end of section \ref{sec8}), or minimize the deviation of the PN ellipsoid from the surface of the Maclaurin ellipsoid (this option is discussed in section \ref{sec11}), or something else. Comparison of the various gauges is facilitated if we operate with the equatorial, $r_e$, and polar, $r_p$, radii of the PN ellipsoid that have been introduced earlier in \eqref{eqra}, \eqref{polra}.

Making use of $r_e$ and $r_p$ we can recast the total mass $M$ in \eqref{totm0} to the form which depends on the linear combination, $E_1+E_2-E_3$, that is 
\be\la{totm1a}
M=\frac{4\pi}{3} \r r_e^2r_p\lt[1-\frac{\kappa}{5}\lt(E_1+E_2-E_3-10{\gimel}_0-\frac{2\o^2}{\pi G\r}\rt)\rt]\;.\ee
The inverse relation will be used to convert the density $\r$ to the total mass,
\be\la{vtx4}
\r=\frac{3M}{4\pi r_e^2r_p}\lt[1+\frac{\kappa}{5}\lt(E_1+E_2-E_3-10{\gimel}_0-\frac{2\o^2}{\pi G\r}\rt)\rt]\;.
\ee 
We shall prove in section \ref{sec8} that the linear combination $E_1+E_2-E_3$ is uniquely defined by the physical equation \eqref{eq1} of the equipotential level surface. Thus, equation (\ref{totm1a}) for the total post-Newtonian mass of the rotating fluid depends on the choice of the free parameters $E_1$ and $E_2$ solely through the equatorial, $r_e$, and polar, $r_p$, radii or, more exactly, on the choice of the ratios: $r_e/a=1+\kappa E_1/2$ and $r_p/b=1+\kappa E_2/2$.

\subsection{Post-Newtonian Angular Momentum}\la{sec7b}

Vector of the post-Newtonian angular momentum, $S^i=(S^x,S^y,S^z)$, is defined by
\citep{fock_1964book,will_1993book,Kopeikin_2011_book}
\be
S^i=S^i_N+\frac1{c^2}S^i_{pN}\;,
\ee
where $S^i_N$ and $S^i_{pN}$ are the Newtonian and post-Newtonian contributions respectively,
\ba\la{spin1a}
S^i_N&=&\int\r({\bm x})\lt({\bm x}\times{\bm v}\rt)^id^3x\;,\\
S^i_{pN}&=&\int\r({\bm x})\lt(v^2+\Pi+6V+\frac{p}{\r}\rt)\lt({\bm x}\times{\bm v}\rt)^id^3x-4\int\r({\bm
x})\lt({\bm x}\times{\bm V}\rt)^id^3x\;,
\ea
and vector-potential ${\bm V}\equiv V^i$ has been given in \eqref{kui8} and \eqref{kui9}.

It can be checked by inspection that in case of axisymmetric mass distribution with a constant density
$\r({\bm x})=\r$, the only non-vanishing component of the angular momentum, is $S^3= S^z\equiv S$. Indeed,
${\bm v}=\{v^i\}=\lt({\bm\omega}\times{\bm x}\rt)^i$, and $\lt({\bm x}\times{\bm v}\rt)^i=\lt({\bm
x}\times\lt({\bm\omega}\times{\bm x}\rt)\rt)^i=\omega^i(x^2+y^2+z^2)-x^i\omega z$. Making use of these relations in \eqref{spin1a} results in
\be\la{gy7}
S_N^x=-\omega\r\int xzd^3x\;,\qquad S_N^y=-\omega\r\int yzd^3x\;,\qquad S_N^z=\omega\r\int (x^2+y^2)d^3x\;.
\ee
Subsequent calculation of the spin components \eqref{gy7} with the help of the spherical coordinates \eqref{newc3} confirms that the two components, $S_N^x=S_N^y=0$, and
$S_N^z\equiv S_N$ where
\be\la{kiv6}
S_N=a^4b\r\omega\int\limits_0^{r(\th)}\int\limits_0^\pi\int\limits_0^{2\pi}r^4\sin^3\th dr d\th d\lambda\;,
\ee
and the boundary of the integration of the radial coordinate, $r(\th)$, is defined in \eqref{ppn8}. After integration in \eqref{kiv6} we obtain,
\be\la{kiv7}
S_N=\frac{8\pi}{15}a^4b\r\omega\lt[1+\frac{\kappa}{14}\lt(24 E_1+ 3E_2+4 E_3  \rt)\rt]\;.
\ee
Replacing the density $\r$ by the total mass $M$ with the help of \eqref{totm0}, makes the Newtonian part of
the angular momentum as follows,
\be\la{ssf5}
S_N=\frac25 Ma^2\omega\lt[1+\frac{\kappa}{35}\lt(32E_1-3E_2+3E_3-70\gimel_0-\frac{14\o^2}{\pi G\r}\rt)\rt]\;.
\ee
The gauge-invariant expression for $S$ can be obtained after making the residual gauge transformation \eqref{tr56} in the defining equation \eqref{spin1a}. Repeating calculations being similar to those which led to the gauge-invariant expression for the mass, we obtain
\be\la{gispin3}
S_N=\frac25 Ma^2\omega\lt[1+\frac{\kappa}{35}\lt(32E'_1-3E'_2+3E'_3-70\gimel_0-\frac{14\o^2}{\pi G\r}-48p+12q+32p\frac{b^2}{a^2}-24q \frac{a^2}{b^2}\rt)\rt]\;.
\ee
The combination of the residual gauge parameters, $-48p+12q+32pb^2/a^2-24qa^2/b^2\not=0$, in general case even if the residual gauge transformation \eqref{tr56} preserves the coordinate volume of integration. This is because in case of spin we integrate over the volume not simply a local mass density $\r$ but the local density of the angular momentum, $\r({\bm x}\times{\bm v})^i$ which is not constant.

It is straightforward to prove that $x$ and $y$ components of $S_{pN}^i$ also vanish due to the axial symmetry,
and only its $z$ component, $S^z_{pN}\equiv S_{pN}$, remains. We notice that
\be
\int\r\lt({\bm x}\times{\bm V}\rt)^zd^3x=\int\r D_1\lt({\bm x}\times{\bm v}\rt)^zd^3x\;,
\ee
where $D_1$ is taken from \eqref{u76}. Therefore, 
\be\la{lin3}
S_{pN}=\int\r\lt(v^2+6V_N+\frac{p}{\r}-4D_1\rt)\lt({\bm x}\times{\bm v}\rt)^zd^3x\;,
\ee
where we have eliminated the compression energy $\Pi$ by including it to the mass density $\r$. 
Making transformation to the coordinates \eqref{newc3} yields
\ba
S_{pN}&=&\omega a^6b\r\int\limits_0^1\int\limits_0^\pi\int\limits_0^{2\pi}\lt\{\omega^2r^2\sin^2\th+7\gimel_0 -
6\gimel_1
- \lt(\gimel_0 + 4 \gimel_1 - 4 \gimel_2\rt) r^2 - 
 2 \lt(3\gimel_0- 15 \gimel_1 + 10 \gimel_2\rt) r^2 \cos^2\th\rt\}r^4\sin^3\th dr d\th d\lambda
 \nonumber\\\la{ssf7}
&=&\frac{4}{35}Ma^2 \omega \lt[2\omega^2+\pi G\r\lt(19{\gimel}_0-16{\gimel}_1\rt)\rt]\;.
\ea
After adding up formulas \eqref{ssf5} and \eqref{ssf7} the total angular momentum becomes
\be\la{totan2}
S=\frac25
Ma^2\omega\lt\{1+\frac{\kappa}{35}\lt[32E_1-3E_2+3E_3+40\lt(3{\gimel}_0-4{\gimel}_1\rt)+\frac{6\o^2}{\pi
G\r}\rt]\rt\}\;.
\ee
Making use of the equatorial radii, $r_e$ defined in \eqref{eqra}, we obtain the final expression for the total
angular momentum
\be\la{angu1}
S=\frac25
Mr_e^2\omega\lt\{1-\frac{\kappa}{35}\Bigl[3(E_1+E_2-E_3)-40\lt(3{\gimel}_0-4{\gimel}_1\rt)-\frac{6\o^2}{\pi
G\r}\Bigr]\rt\}\;.
\ee 
This expression depends only on the linear combination of the parameters, $E_1+E_2-E_3$, both explicitly and implicitly (through the mass $M$ in equation \eqref{totm1a}), which is uniquely fixed in the chosen coordinate system by the equation of the level surface \eqref{eq1}. The total angular momentum $S$ 
depends explicitly only on the parameter $E_1$ through the equatorial radius $r_e$. It depends implicitly on the parameters $E_1$ and $E_2$ through the mass $M$. The two parameters $E_1$ and $E_2$ can be chosen arbitrary depending on our preferences and the goals which we want to reach in relativistic geodesy.

\section{Post-Newtonian Equation of the level surface}\la{sec8}

The figure of the rotating fluid body is defined by the boundary condition of vanishing pressure, $p=0$. The
boundary surface, $p=0$, is called the level surface. Relativistic Euler equation derived for the rigidly rotating fluid
body, tells us \citep{Kopejkin_1991,kop_2015PhLA} that the level surface coincides with the equipotential surface of the post-Newtonian
gravitational potential $W$ which is given by \citep{Kopeikin_2011_book}
\begin{equation}\la{lsur}
W=\frac12\omega^2\s^2+V_N+\kappa
V_{pN}+\frac1{c^2}\lt(\frac18\omega^4\s^4+\frac32\omega^2\s^2V_N-4\omega^2\s^2D_1-\frac12 V_N^2+\Phi\rt)\;,
\end{equation}
where $\kappa\equiv\pi G\r a^2/c^2$, and the potentials $V_N, V_{pN}, D_1,\Phi$ have been explained in sections
\ref{sec5} and \ref{sec6}. After substituting these potentials to equation \eqref{lsur} it can be presented as
a quadratic polynomial with respect to the function $C({\bm x})$,
\be\la{w45}
W({\bm x})=W_0+W_1C({\bm x})+W_2C^2({\bm x})\;,
\ee
where the coefficients of the expansion are polynomials of the $z$ coordinate only.
In particular, the coefficient $W_0$ is a polynomial of the fourth order,
\be\la{sur1}
W_0=K_0+K_1\frac{z^2}{b^2}+K_2\frac{ z^4}{b^4}\;,
\ee
where 
\ba
K_0&=&\frac12\omega^2a^2+\pi G\r a^2\lt({\gimel}_0-{\gimel}_1\rt)\\\nonumber
&&+\frac1{8c^2}\omega^4a^4+\frac12\kappa
\omega^2a^2\lt(5\gimel_0-17\gimel_1+18\gimel_2-6\gimel_3\rt)\\\nonumber
&&+
\frac12\kappa\pi G\r a^2\lt[6\gimel_0^2-\gimel_0 (22 \gimel_1 - 15 \gimel_2 + 6 \gimel_3) + \gimel_1 (29
\gimel_1 - 40 \gimel_2 + 18 \gimel_3)\rt]\\\nonumber
&&+\kappa\pi G\r
a^2\Biggl[\lt(\gimel_0-4\gimel_1+10\gimel_2-12\gimel_3+6\gimel_4\rt)E_1+2\lt(\gimel_1-4\gimel_2+6\gimel_3-3\gimel_4\rt)E_3+
6\lt(\gimel_2-2\gimel_3+\gimel_4\rt)E_2  \Biggr]
\;,\\
K_1&=&-\frac12\omega^2a^2-\pi G\r a^2\lt({\gimel}_0-3{\gimel}_1\rt)\\\nonumber
&&-\frac1{4c^2}\omega^4a^4-\kappa\omega^2 a^2\lt(5\gimel_0-40\gimel_1+66\gimel_2-30\gimel_3\rt)-
\kappa\pi G\r a^2\lt(6\gimel_0^2-56 \gimel_0 \gimel_1 + 99 \gimel_1^2 + 67 \gimel_0 \gimel_2 - 176 \gimel_1
\gimel_2 - 30 \gimel_0 \gimel_3 + 90 \gimel_1 \gimel_3\rt)\\\nonumber
&&-\kappa\pi G\r
a^2\Biggl[2\lt(\gimel_0-12\gimel_1+42\gimel_2-60\gimel_3+30\gimel_4\rt)E_1-\lt(\gimel_0-18\gimel_1+78\gimel_2-120\gimel_3+60\gimel_4\rt)E_3-
12\lt(\gimel_1-6\gimel_2+10\gimel_3-5\gimel_4\rt)E_2  \Biggr]\;,\\
K_2&=&\frac1{8c^2}\omega^4a^4+\frac12\kappa\omega^2
a^2\lt(5\gimel_0-63\gimel_1+130\gimel_2-70\gimel_3\rt)\\\nonumber
&&+\frac16\kappa\pi G\r a^2\lt[18 \gimel_0^2 - 5 \gimel_0 (54 \gimel_1 - 89 \gimel_2 + 42 \gimel_3) + 
 \gimel_1 (543 \gimel_1 - 1160 \gimel_2 + 630 \gimel_3)\rt]\\\nonumber
&&+\kappa\pi G\r a^2\lt(\gimel_0-20\gimel_1+90 {\gimel}_2 -140 {\gimel}_3+70\gimel_4 \rt)\lt(E_1+E_2-E_3\rt)
 \;.
\end{eqnarray}
The coefficient $W_1$ in \eqref{w45} is a polynomial of the second order,
\be\la{w46}
W_1=P+P_1\frac{z^2}{b^2}\;,
\ee
where
\ba\la{w47}
P&=&\frac12\omega^2a^2\lt[1+\frac1{2c^2}\omega^2a^2+\kappa\lt(3\gimel_0-18\gimel_1+28\gimel_2-12\gimel_3\rt)\rt]-\pi
G\rho a^2\gimel_1\\\nonumber
&&-\kappa\pi G\rho a^2\Biggl[2\lt( \gimel_1- 5 \gimel_2+9\gimel_3-6\gimel_4\rt) E_1 - \lt(\gimel_1 - 8
\gimel_2+18\gimel_3-12\gimel_4\rt) E_3 - 6 \lt(\gimel_2-3\gimel_3+2\gimel_4\rt) E_2\Biggr]\\\nonumber
&&-\kappa\pi G\rho a^2\lt(6\gimel_0 \gimel_1 - 11 \gimel_1^2-11\gimel_0\gimel_2 +28\gimel_1
\gimel_2+6\gimel_0\gimel_3-18\gimel_1\gimel_3\rt)\;,\\
P_1&=&-\frac1{4c^2}\omega^2a^4-\frac32\kappa\omega^2a^2\lt(\gimel_0-16\gimel_1+36\gimel_2-20\gimel_3\rt)\\\nonumber
&&+\kappa\pi G\rho a^2\lt[2\lt(3 \gimel_1 - 25 \gimel_2  + 51 \gimel_3 - 30 \gimel_4\rt)\lt(E_1+E_2 - E_3\rt)
+20\gimel_0 \gimel_1 -45 \gimel_1^2-55\gimel_0 \gimel_2 +140 \gimel_1 \gimel_2+30\gimel_0 \gimel_3-90\gimel_1
\gimel_3 \rt] \;.
\ea
The coefficient $W_2$ in \eqref{w45} is constant,
\ba\la{w78}
W_2&=&\frac1{8c^2}\omega^4a^4-\frac12\kappa\omega^2a^2\lt(3\gimel_1-2\gimel_2-6\gimel_3\rt)\\\nonumber
&&+\kappa\pi G\rho
a^2\Biggl[\lt(\gimel_2-6\gimel_3+6\gimel_4\rt)\lt(E_1+E_2-E_3\rt)-\frac12\lt(\gimel_1^2+\gimel_0\gimel_2-8\gimel_1\gimel_2-6\gimel_0\gimel_3+18\gimel_1\gimel_3\rt)\Biggr]\;.
 \ea

Let us recall that the coordinates on the surface of the PN ellipsoid are denoted as $\bar x$, $\bar y$, $\bar z $.
On the level surface of the PN ellipsoid we have all three coordinates interconnected by
equation \eqref{uui8} of the PN ellipsoid, $C({\bar{\bm x}})=\kappa\alpha_0(\bar z)$, so that \eqref{w45}
becomes
\be
\bar W\equiv \bar W_0+\kappa \bar W_1\alpha_0(\bar z)\;,
\ee
and the term with $W_2\sim{\cal O}\lt(\kappa^2\rt)$, is discarded as negligibly small.
After reducing similar terms, the potential $\bar W$ on the level surface is simplified to the polynomial of
the fourth order,
\be
\bar W=K'_0+K'_1\frac{\bar z^2}{b^2}+K'_2\frac{\bar z^4}{b^4}\;,
\ee
where 
\ba
K'_0&=&K_0+\kappa\lt(\frac12\omega^2a^2-\pi G\r a^2 \gimel_1\rt)E_1\;,\\
K'_1&=&K_1+\kappa\lt(\frac12\omega^2a^2-\pi G\r a^2 \gimel_1\rt)\lt(E_3-2E_1\rt)\;,\\
K'_2&=&K_2+\kappa\lt(\frac12\omega^2a^2-\pi G\r a^2 \gimel_1\rt)\lt(E_1+E_2-E_3\rt)\;.
\ea
Because the potential $\bar W$ is to be constant on the level surface \citep{kop_2015PhLA}, the numerical
coefficients $K'_1$ and $K'_2$ must vanish. The first condition, $K'_1=0$, yields a relation between the the
angular velocity of rotation, $\omega$, and oblateness, $e$, of the rotating fluid body,
\ba\la{pm6y}
&&\frac{\omega^2}{2\pi
G\r}\lt[1+\frac{\omega^2a^2}{2c^2}+2\kappa\lt(5\gimel_0-40\gimel_1+66\gimel_2-30\gimel_3-\frac12E_3+E_1\rt)\rt]=3{\gimel}_1-{\gimel}_0\\\nonumber
&&\quad - \kappa\lt(6\gimel_0^2-56\gimel_0
{\gimel}_1+99\gimel_1^2+67\gimel_0\gimel_2-176\gimel_1\gimel_2-30\gimel_0\gimel_3+90\gimel_1\gimel_3\rt)-\kappa\lt(\gimel_0-7\gimel_1+6\gimel_2\rt)\lt(E_1-E_2\rt)\\\nonumber
&&\quad-\kappa\lt(\gimel_0-19\gimel_1+78\gimel_2-120\gimel_3+60\gimel_4\rt)\lt(E_1+E_2-E_3\rt)\;.
\ea
Equation \eqref{pm6y} generalizes the famous result that was first obtained by Colin Maclaurin in 1742, from
the Newtonian theory of gravity to the realm of general relativity. Physical meaning of the post-Newtonian Maclaurin relation is that it connects four parameters of the rotating ellipsoid made up of a homogeneous fluid -- its eccentricity $e$, the semimajor axis $a$, the angular velocity of revolution $\o$, and density $\r$. In the Newtonian case the Maclaurin relation connects only three parameters: $e,\o,\r$. It also gives a rigorous mathematical proof of Newton's original claim that a rotating body must oblate in the direction of rotational axis \citep{Chandr_1967a}.  

Equation \eqref{pm6y} can be further simplified by replacing the eccentricity $e$ of the Maclaurin ellipsoid with that $\e$ of the PN ellipsoid by making use of \eqref{obt5}. We introduce functions $\gimel_0(\e)$ and $\gimel_1(\e)$ that are given by equations \eqref{inta0} and \eqref{inta0} after the formal replacement of $e$ in those equations with $\e$. We expand $\gimel_0(\e)$ and $\gimel_1(\e)$ in the Taylor series with respect to the relativistic parameter $\kappa$, and find out that
\ba\la{vv12}
\gimel_0(\e)&=&\gimel_0+\kappa\lt(E_1-E_2\rt)\gimel_1+{\cal O}\lt(\kappa^2\rt)\;\\\la{vv13}
\gimel_1(\e)&=&\gimel_1+\kappa\frac{E_1-E_2}{e^2}\lt(3\gimel_1-\gimel_0\rt)+{\cal O}\lt(\kappa^2\rt)\;,
\ea
where $\gimel_0\equiv\gimel_0(e)$ and $\gimel_1\equiv\gimel_1(e)$ are given by equations \eqref{inta0} and \eqref{inta0}. We express $\gimel_0$ and $\gimel_1$ in terms of $\gimel_0(\e)$ and $\gimel_1(\e)$ by inverting \eqref{vv12}, \eqref{vv13}, and then, substitute the expressions having been  obtained, to the Newtonian part, $3\gimel_1-\gimel_0$, in the right side of \eqref{pm6y}. It turns out that all terms depending explicitly on $E_1$ and $E_2$, cancel each other mutually so that \eqref{pm6y} takes on a more elegant form
\ba\la{pm6y1}
&&\frac{\omega^2}{2\pi
G\r}\lt[1+\frac{\omega^2a^2}{2c^2}+2\kappa\lt(5\gimel_0-40\gimel_1+66\gimel_2-30\gimel_3\rt)\rt]=3{\gimel}_1(\e)-{\gimel}_0(\e)\\\nonumber
&&\quad - \kappa\lt(6\gimel_0^2-56\gimel_0
{\gimel}_1+99\gimel_1^2+67\gimel_0\gimel_2-176\gimel_1\gimel_2-30\gimel_0\gimel_3+90\gimel_1\gimel_3\rt)\\\nonumber
&&\quad-\kappa\lt(\frac{\omega^2}{2\pi
G\r}+\gimel_0-19\gimel_1+78\gimel_2-120\gimel_3+60\gimel_4\rt)\lt(E_1+E_2-E_3\rt)\;.
\ea
We can see that the Maclaurin relation in the form of \eqref{pm6y1} depends explicitly only on the linear combination of the coefficients, $E_1+E_2-E_3$, which is fixed by the equation of the level surface \eqref{eq1}. Dependence on the free parameters $E_1$ and $E_2$ enters \eqref{pm6y1} only through the eccentricity $\epsilon$ of the PN ellipsoid.
 
The second condition, $K'_2=0$, yields an algebraic equation for the linear combination of three coefficients $E_1+E_2-E_3$, namely
\ba\la{eq1}
&&\lt(\frac{\omega^2}{2\pi G\r}+\gimel_0-21\gimel_1+90 {\gimel}_2 -140 {\gimel}_3+70\gimel_4
\rt)\lt(E_1+E_2-E_3\rt)=\\\nonumber
&&-\frac{\omega^4}{8\pi^2 G^2\r^2}-\frac{\omega^2}{4\pi
G\r}\lt(5\gimel_0-63\gimel_1+130\gimel_2-70\gimel_3\rt)\\\nonumber
&&-\frac16\lt[18 \gimel_0^2 - 5 \gimel_0 (54 \gimel_1 - 89 \gimel_2 + 42 \gimel_3) + 
 \gimel_1 (543 \gimel_1 - 1160 \gimel_2 + 630 \gimel_3)\rt]\;.
\ea
Equation \eqref{eq1} imposes one physical constraint on the coefficients $E_1,E_2,E_3$ defining the shape of the
PN ellipsoid \eqref{mnh763}. Two other algebraic equations are required to fix the numerical coefficients $E_1$ and $E_2$.
Because of the residual gauge freedom, explained above in section \ref{sec4}, the two equations can be chosen
arbitrary. This property of the gauge freedom of the post-Newtonian theory of figures of rotating fluid bodies has been noticed by Chandrasekhar \citep{Chandr_1965ApJ142_1513,Chandr_1967ApJ147_334} who limited the gauge freedom by imposing one condition of the conservation of the volume element of the fluid under the gauge transformation \eqref{po98}--\eqref{po23}. Chandrasekhar believed that the second condition remains free and can be chosen arbitrary.    

On the other hand, Bardeen \citep{Bardeen_1971ApJ} pointed out that there exist two gauge-fixing conditions arising naturally from the astrophysical point of view and specifying uniquely the shape of the uniformly rotating fluid in the post-Newtonian approximation. Namely, he suggested to equate the total mass and angular momentum of the (Newtonian) Maclaurin ellipsoid to those of the post-Newtonian ellipsoid \citep{footnote-2} under condition that they both have equal mass density $\r$. Later on,
Bardeen's conditions were accepted and implemented by Chandrasekhar \citep{Chandr_1971ApJ167_455} and Pyragas et al \citep{Pyragas1974Ap&SS27_453} as well. We can easily impose the Bardeen gauge by making use of our equations \eqref{totm0} and \eqref{totan2}. The total mass and the angular momentum of the Maclaurin ellipsoid are given by equations
\be\la{nhb47}
M_{\rm Maclaurin}=\frac{4\pi}3\r a^2b\;,\qquad\qquad S_{\rm Maclaurin}=\frac25 M_{\rm Maclaurin}a^2\omega\;.
\ee
Equating $M=M_{\rm Maclaurin}$ in \eqref{totm0}, and $S=S_{\rm Maclaurin}$ in \eqref{totan2}, yield the Bardeen gauge conditions
\ba\la{con1a}
8E_1+3E_2+2E_3&=&-20{\gimel}_0-\frac{4\o^2}{\pi G\r}\;,\\
\la{con2a}
32E_1-3E_2+3E_3&=&-40\lt(3{\gimel}_0-4{\gimel}_1\rt)-\frac{6\o^2}{\pi G\r}\;.
\ea
It should be emphasized, however, that these constraints suggest that neither equatorial, $a$, nor polar radii, $b$ of the (Newtonian) Maclaurin ellipsoid are equal to the equatorial, $r_e$, and the polar, $r_p$, radii of the PN ellipsoid respectively: $r_e\not=a$, $r_p\not=b$ as shown in the left panel of Fig. \ref{fig1a}.
Indeed, assuming that $r_e=a$, $r_p=b$ imposes two constraints on the parameters $E_1$ and $E_2$ which are simply $E_1=E_2=0$ as follows from \eqref{eqra} and \eqref{polra}. This corresponds to the geometrical shape of the PN ellipsoid shown in the right panel of Fig. \ref{fig1a}. 
However, these two constraints makes three equations \eqref{eq1}, \eqref{con1a}, \eqref{con2a} for the remaining parameter $E_3$ incompatible with each other. Thus, in geodetic applications of the relativistic theory of rotating fluids we have to decide which option has more practical advantages: 1) to keep the same mass and angular momenta but different axes of the Maclaurin and PN ellipsoids, or 2) to keep their axes equal but to abandon the equality of their masses and angular momenta. 

In order to understand better the physical meaning of the Bardeen-Chandrasekhar gauge, let us consider three masses introduced earlier: 1) the Newtonian mass $M_{\rm Maclaurin}$ of the Maclaurin ellipsoid \eqref{nhb47}, 2) the Newtonian mass $M_N$ of the PN ellipsoid \eqref{lov6}, and 3) the total post-Newtonian mass $M$ of the PN ellipsoid \eqref{totm0}. The coordinate volumes of the Maclaurin and PN ellipsoids are different while the density $\r$ of the constituting matter is the same. Hence, $M_{\rm Maclaurin}$ cannot be equal to $M_N$ unless the trivial case, $E_1=E_2=E_3=0$, that is when the surface PN ellipsoid coincides with that of the Maclaurin ellipsoid. The coordinate volumes occupied by the Newtonian mass $M_N$, and the post-Newtonian mass $M$ of the PN ellipsoid, are the same. Nonetheless, $M\not=M_N$ because $M_N$ is solely comprised of the rest mass of baryons while the total post-Newtonian mass $M$ includes additional (positive) contribution $M_{pN}$ given in \eqref{pnm123} and corresponding to the internal kinetic, compressional and gravitational energy of the body's matter. By changing the numerical values of the freely adjustable parameters $E_1$ and $E_2$ we can decrease the baryon mass $M_N$ by the amount that compensates the (positive) post-Newtonian contribution $M_{pN}$, and make $M_{\rm Maclaurin}$ equal to $M$. This is achieved under condition that the first gauge-fixing equation \eqref{con1a} is satisfied. The same reasoning is valid with regard to the comparison of the Newtonian and post-Newtonian angular momenta which leads to the second gauge-fixing condition \eqref{con2a}.  

It should be understood that the Bardeen-Chandrasekhar gauge imposed on the coordinates to build the post-Newtonian metric of a rotating fluid planet or a star, is not the only possible one in the most general case. It is convenient in astrophysics because it facilitates unambiguous comparison of various physical properties of the rotating Newtonian configurations with respect to relativistic stars having the same mass and angular momenta which are the integrals of the equations of motion. However, the primary goal of geodesy differs from astrophysics and is to build the terrestrial reference frame that is the most precise and adequate for interpretation of measurements of baseline's length, motion of geodetic stations, deflections of the plumb line and variations (anomalies) of Earth's gravitational field. Until recently these type of measurements have been referred to a homogeneous reference ellipsoid possessing a rather simple and exact analytic description of the normal gravitational field of the Earth. It seems reasonable to make the post-Newtonian reference configuration in relativistic geodesy as close to the Maclaurin reference ellipsoid as possible to minimize the contribution of relativistic corrections to the coordinates and velocities of the geodetic stations. This can be achieved with the choice of the coefficients $E_1=E_2=0$ in equation \eqref{1} of the PN ellipsoid which is not the Bardeen-Chandrasekhar gauge. We continue discussion of this question in section \ref{sec11}.

Besides making a decision which gauge is the most appropriate in relativistic geodesy, we have to establish mathematical relations between the parameters of the relativistic PN ellipsoid and the gravimetric measurements of the Earth's gravity force on its topographic surface. These relations are known in classic (Newtonian) geodesy as the theorems of Pizetti and Clairaut \citep{Torge_1989gravbook}, and they connect parameters of the Maclaurin ellipsoid, namely, the semimajor and semiminor axes $a$ and $b$, mass $M$ and the angular velocity of rotation $\o$, with the physically measured values of the
gravity force at the pole and equator of the ellipsoid \citep{pizz_1913,Torge_2012_book}. 

Let us denote the Newtonian force of gravity by $\g^N_i({\bm x})=\{\g^N_x,\g^N_y,\g^N_z\}$, the force of gravity measured at the pole of the ellipsoid by $\g^N_p\equiv\g^N_z(x=0,y=0,z=b)$, and the force of gravity measured 
at equator by $\g^N_e\equiv\g^N_y(x=0,y=a,z=0)$ \citep{footnote-3}. Due to the rotational symmetry of the ellipsoid the equatorial point can be, in fact, chosen arbitrary.
The classic form of the theorem of Pizetti is \citep[eq. (4.42)]{Torge_2012_book}
\be\la{pth56}
2\frac{\g^N_e}{a}+\frac{\g^N_p}{b}=\frac{3 G M_N}{a^2b} - 2\o^2\;,
\ee 
while the theorem of Clairaut states \citep[eq. (4.43)]{Torge_2012_book}
\be\la{clr29}
\frac{\g^N_e}{a}-\frac{\g^N_p}{b}=\frac{3 G
M_N}{2a^2b}\frac{3e-e^3-3\sqrt{1-e^2}\arcsin
e}{e^3}+\o^2\;.
\ee
The main value of the theorems of Pizetti and Clairaut is that they allow us to calculate explicitly the normal gravity field of the reference level ellipsoid in terms of only four ellipsoid's parameters -- $M_N,\o, a, b$ -- in agreement with the Stokes-Poincar\'e theorem \citep{footnote-4}.

Theorems \eqref{pth56} and \eqref{clr29} were crucial in geodesy of XIX-th century because they helped scientists to realize that the geometric shape of Earth's figure can be determined not only from the geometric measurements of the geodetic arcs but, independently, by rendering the intrinsic measurements of the force of gravity on Earth's surface \citep{Zund_1994_book}. The gravity-geometry correspondence expressed in the form of the two theorems \eqref{pth56} and \eqref{clr29}, led a number of scientists from Lobachevsky to Einstein to a gradual understanding that the gravity force and geometry of curved spacetime must be interrelated. This geodesy-inspired way of thinking culminated in XX-th century in the development of general theory of relativity by A. Einstein. 

We derive the post-Newtonian analogues of the Pizzetti and Clairaut theorems in the next two sections. We show that the parameters entering the post-Newtonian formulation of these theorems are still the same four parameters as in the Newtonian approximation with a corresponding replacement of the Newtonian values of the parameters by their relativistic counterparts.  

\section{Post-Newtonian Theorem of Pizzetti}\la{sec9}

We denote the post-Newtonian force of gravity $\g_i({\bm x})=\{\g_x,\g_y,\g_z\}$. The force of gravity measured by a local observer on the equipotential surface of the Earth gravity field has been derived in \citep{Kopejkin_1991,kop_2015PhLA} and is given by equation
\citep{Kopeikin_2011_book}
\be\la{fg56}
\gamma_i= \lt[\Lambda^j{}_i\pd_j W\rt]_{{\bm x}=\bar{\bm x}}\;,
\ee
where $\pd_i\equiv\pd/\pd x^i$, the post-Newtonian gravity potential $W$ has been defined in \eqref{w45},  
\be
\Lambda^j{}_i=\delta^{ij}\lt(1-\frac1{c^2}V_N\rt)-\frac{1}{2c^2}v^iv^j\;,
\ee
is the matrix of transformation from the global (GCRS) coordinates to the local inertial (topocentric) coordinates of observer,
$v^i=\lt({\bm\omega}\times{\bm x}\rt)^i$ is velocity of the observer with respect to the global coordinates,
and $V_N$ is the Newtonian potential \eqref{nn1}. It is worth emphasizing that we, first, take the partial derivative in
\eqref{fg56}, and then, take the spatial coordinates, ${\bm x}$, on the equipotential surface, ${\bm x}\rightarrow\bar{\bm x}$. 

Velocity $v^i=({\bm\o}\times{\bm x})^i$ is orthogonal to the gradient $\pd_i W$ everywhere, that is 
\be\la{kk9}
v^i\pd_i W=0\;.
\ee
Indeed, it is easy to prove that
\be\la{kjh}
v^i\pd_i W=\o\lt(x\pd_yW-y\pd_xW\rt)\;.
\ee
Partial derivatives of $W$ are calculated from \eqref{w45},
\be\la{nev2}
\pd_xW=\frac{dW}{dC}\pd_x C({\bm x})=\frac{dW}{dC}\frac{2x}{a^2}\;,\qquad \pd_yW=\frac{dW}{dC}\pd_y C({\bm
x})=\frac{dW}{dC}\frac{2y}{a^2}\;.
\ee
Substituting the partial derivatives from \eqref{nev2} to \eqref{kjh} yields \eqref{kk9} which was to be demonstrated. 
After accounting for \eqref{kk9}, equation \eqref{fg56} is simplified to
\be\la{fg57}
\gamma_i(\bar{\bm x})=\lt[\lt(1-\frac1{c^2}V_N\rt)\pd_iW\rt]_{{\bm x}=\bar{\bm x}}\;.
\ee

We take the PN ellipsoid \eqref{1} as the equipotential surface enclosing the entire rotating mass and
denote the post-Newtonian force of gravity on the pole by $\g_p\equiv\g_z(x=0,y=0,z=r_p)$ and the force of gravity
on the equator by $\g_e\equiv\g_y(x=0,y=r_e,z=0)$ with the equatorial $r_e$ and polar $r_p$ radii defined in (\ref{eqra}) and (\ref{polra}) respectively. Taking the partial derivative from $W$ in \eqref{fg57} yields
\ba\la{fg58}
\g_p&=&\frac{2\pi G\r
a^2}{b}\lt(\gimel_0-2\gimel_1\rt)+16\frac{\omega^2a^2}{b}\kappa\lt(\gimel_1-3\gimel_2+2\gimel_3\rt)\\\nonumber
&&+\frac{\pi G\r
a^2}{b}\kappa\lt[2\lt(\gimel_0-5\gimel_1+4\gimel_2\rt)E_1-\lt(\gimel_0-8\gimel_1+8\gimel_2\rt)E_2
-2\lt(\gimel_0-17\gimel_1+60\gimel_2-76\gimel_3+32\gimel_4\rt)\lt(E_1+E_2-E_3\rt)\rt]\\\nonumber
&&+\frac{4\pi G\r
a^2}{3b}\kappa\lt[\gimel_0\lt(27\gimel_1-56\gimel_2+24\gimel_3\rt)-2\gimel_1\lt(33\gimel_1-74\gimel_2+36\gimel_3\rt)\rt]\;,\\
\la{fg58gt}
\g_e&=&a \lt(2 \gimel_1 G \pi \rho - \omega^2\rt)-\frac{\omega^4a^3}{2c^2}+\kappa\lt[-3 \gimel_0 + 18 \gimel_1
- 28 \gimel_2 + 12 \gimel_3 + \lt( \gimel_0 - \gimel_1 -\frac12 E_1\rt) \rt]\omega^2 a\\\nonumber
&&+\kappa \pi G\r a\lt[\lt(3\gimel_1-4\gimel_2\rt)E_1 - 2\lt(\gimel_1-2\gimel_2\rt)E_2 + 2\lt(\gimel_1-8
\gimel_2+18 \gimel_3 - 12 \gimel_4 \rt )\lt(E_1+E_2-E_3\rt)\rt]\\\nonumber
&&+2\kappa \pi G\r
a\lt[\gimel_0\lt(5\gimel_1-11\gimel_2+6\gimel_3\rt)-2\gimel_1\lt(5\gimel_1-14\gimel_2+9\gimel_3\rt)\rt]\;.
\ea

The right side of \eqref{fg58}, \eqref{fg58gt} depends on the semimajor and semimainor axes of the Maclaurin ellipsoid, $a$ and $b$, but they are not defining parameters of the PN ellipsoid which are the equatorial and polar radii, $r_e$ and $r_p$,  given in \eqref{eqra}, \eqref{polra} respectively.
Moreover, the right side of \eqref{fg58}, \eqref{fg58gt} depends on the gauge parameters $E_1$, $E_2$. We replace parameters $a$ and $b$ with $r_e$ and $r_p$, and form a linear combination generalizing the Newtonian theorem of Pizetti to the post-Newtonian approximation,
\ba\la{fgp}
2\frac{\g_e}{r_e}+\frac{\g_p}{r_p}&=&2\pi G\r\lt[2 \gimel_1 + \frac{a^2}{b^2} (\gimel_0 - 2 \gimel_1)\rt] - 2
\omega^2-\frac{1}{c^2}\omega^4a^2\\\nonumber
&-&2\kappa\lt[2 \gimel_0 - 17 \gimel_1 +28 \gimel_2 - 12 \gimel_3 - 8\frac{a^2}{b^2}\lt( \gimel_1 -3\gimel_2 +2
\gimel_3\rt)\rt]\omega^2\\\nonumber
&+&2\kappa \pi G\r\lt[2 \gimel_1 - 16 \gimel_2 + 36 \gimel_3 - 24 \gimel_4-\frac{a^2}{b^2}\lt(\gimel_0 - 17
\gimel_1 + 60 \gimel_2 - 76 \gimel_3 + 32 \gimel_4\rt)\rt]\lt(E_1+E_2-E_3\rt)\\\nonumber
&&+2\kappa \pi
G\r\lt[\lt(2\gimel_1-4\gimel_2\rt)+\frac{a^2}{b^2}\lt(\gimel_0-5\gimel_1+4\gimel_2\rt)\rt]\lt(E_1-E_2\rt)\\\nonumber
&&+4\kappa \pi G\r\lt[\gimel_0 \lt(5 \gimel_1 - 11 \gimel_2 + 6 \gimel_3\rt) - 2 \gimel_1 \lt(5 \gimel_1 - 14
\gimel_2 + 9 \gimel_3\rt)\rt]\\\nonumber
&&+\frac{4a^2}{3b^2}\kappa \pi G\r\lt[\gimel_0 (27 \gimel_1 - 56 \gimel_2 + 24 \gimel_3) - 2 \gimel_1 (33
\gimel_1 - 74 \gimel_2 + 36 \gimel_3)\rt]\;.
\ea
It seems that the right side of \eqref{fgp} still depends explicitly on the gauge parameters $E_1, E_2$. However, 
making use of integrals given in appendix \ref{appen1}, we can check that the numerical coefficient standing in
front of the difference, $E_1-E_2$, vanishes identically, so that \eqref{fgp} is simplified to
\ba\la{fgp1}
2\frac{\g_e}{r_e}+\frac{\g_p}{r_p}&=&4\pi G\r - 2 \omega^2-\frac{1}{c^2}\omega^4a^2\\\nonumber
&-&\frac{\kappa}{3e^7}\lt[e(1-e^2)(105 - 104 e^2 + 42 e^4)- 
 3 \sqrt{1 - e^2} (5-4e^2)(7 -6 e^2 +2 e^4) \arcsin e\rt]\omega^2\\\nonumber
&+&\frac{\kappa \pi G\r}{12e^9} \lt(7 - 4 e^2\rt)\lt[  5e(21 - 31 e^2 + 10 e^4) - 
   3\sqrt{1 - e^2} (35 - 40 e^2 + 8 e^4) \arcsin e\rt]\lt(E_1+E_2-E_3\rt)\\\nonumber
&+&\frac{\kappa \pi G\r}{3e^{10}}\lt(1-e^2\rt)
\lt[-315 e^2 + 621 e^4 - 250 e^6 + 24 e^8 + 
 2 e \sqrt{1 - e^2} (315 - 516 e^2 + 169 e^4 - 18 e^6) \arcsin e\rt.\\\nonumber&&\phantom{+++++++}\;\lt. - 
 3 (105 - 242 e^2 + 178 e^4 - 40 e^6) \arcsin^2 e\rt]\;.
\ea
Relation \eqref{fgp1} depends only on the linear combination $E_1+E_2-E_3$ of the parameters which has been already fixed by equation \eqref{eq1} of the level surface, and can be expressed solely as a function of the eccentricity $e$. 

In order to compare \eqref{fgp1} with its classic counterpart \eqref{pth56}, we convert the constant density $\r$, to the total relativistic mass $M$ of the PN ellipsoid by making use of \eqref{vtx4}. It recasts \eqref{fgp1} to
\ba\la{pipi1}
2\frac{\g_e}{r_e}+\frac{\g_p}{r_p}&=&\frac{3 G M}{r_e^2r_p} - 2\o^2-\frac{\o^4r_e^2}{c^2}\\\nonumber
&+&\frac{3\sqrt{1-\epsilon^2}}{16\epsilon^9r_e^2r_p^2}\frac{G^2M^2}{c^2}\lt[\epsilon\sqrt{1-\epsilon^2}(-315 +621 \epsilon^2 -250 \epsilon^4 +24 \epsilon^6) + 
 2 (315 - 831 \epsilon^2 + 685 \epsilon^4 - 187 \epsilon^6+42\epsilon^8)\arcsin \epsilon\rt]\\\nonumber
&+&\frac{1}{20\epsilon^7r_p}\frac{GM\o^2}{c^2}\lt[\epsilon(-525 +1045 \epsilon^2 -730 \epsilon^4 +234 \epsilon^6) + 
 15 \frac{35 - 93 \epsilon^2 + 92 \epsilon^4 - 42 \epsilon^6+8\epsilon^8}{\sqrt{1-\epsilon^2}}\arcsin \epsilon\rt]\\\nonumber
&-&\frac{9}{16}\frac{G^2M^2}{c^2}\frac{105 - 347 \epsilon^2 + 420 \epsilon^4 - 218 \epsilon^6 + 40 \epsilon^8}{\epsilon^{10}r_e^2r_p^2}\arcsin^2\epsilon+\frac{3}{320\epsilon^8r_e^2r_p^2}\frac{ G^2M^2}{c^2}\lt(E_1+E_2-E_3\rt)
\\\nonumber
&\times&\lt[ 3675 - 7525 \epsilon^2 +4850 \epsilon^4-1000\epsilon^6-48\epsilon^8 - 15 \frac{\sqrt{1 -
\epsilon^2}}{\epsilon} (7 - 4 \epsilon^2)(35 - 40 \epsilon^2 + 8 \epsilon^4) \arcsin \epsilon\rt]\;,
\ea
where we have used in the post-Newtonian terms the eccentricity $\epsilon$ defined in \eqref{im76} instead of $e$  because they differ only in the post-Newtonian terms, and assume that the combination of the parameters $E_1+E_2-E_3=f(\epsilon)$ with function $f(\epsilon)$ given by formula \eqref{eq1}.

Equation \eqref{pipi1} represents the Pizzetti theorem generalizing the classic result (\ref{pth56}) to the domain of the post-Newtonian approximation. It tells us that similarly to the Newtonian theory, the linear combination of the post-Newtonian forces of gravity measured on the surface of the PN ellipsoid at the pole and equator, is a function of only four parameters -- the post-Newtonian mass $M$, the angular velocity of rotation $\o$, and the equatorial and polar radii, $r_e$ and $r_p$, of the PN ellipsoid.

\section{Post-Newtonian Theorem of Clairaut}\la{sec10} 
  
In order to derive the post-Newtonian analogue of the Clairaut theorem \eqref{clr29} we follow its classic derivation given, for example, in \citep{pizz_1913}. To this end we subtract the ratio of the force of gravity \eqref{fg58} measured at the pole to $r_p$ from that of the force of gravity \eqref{fg58gt} measured on equator to $r_e$. We get
\ba\la{mec8}
\frac{\g_e}{r_e}-\frac{\g_p}{r_p}&=&2\pi G\r\lt[- \gimel_1 + \frac{a^2}{b^2} (\gimel_0 - 2 \gimel_1)\rt] +
\omega^2+\frac{1}{2c^2}\omega^4a^2\\\nonumber
&&+\kappa\lt[ 2\gimel_0 - 17 \gimel_1 +28 \gimel_2 - 12 \gimel_3 +16 \frac{a^2}{b^2}\lt( \gimel_1 -3 \gimel_2
+2 \gimel_3\rt)\rt]\omega^2\\\nonumber
&&+\kappa \pi G\r\lt[-2 \gimel_1 +16 \gimel_2 - 36 \gimel_3 + 24 \gimel_4-\frac{2a^2}{b^2}\lt(\gimel_0 - 17
\gimel_1 + 60 \gimel_2 - 76 \gimel_3 + 32 \gimel_4\rt)\rt]\lt(E_1+E_2-E_3\rt)\\\nonumber
&&-2\kappa \pi
G\r\lt[\lt(\gimel_1-2\gimel_2\rt)-\frac{a^2}{b^2}\lt(\gimel_0-5\gimel_1+4\gimel_2\rt)\rt]\lt(E_1-E_2\rt)\\\nonumber
&&+2\kappa \pi G\r\lt[ \gimel_0 (-5 \gimel_1 + 11 \gimel_2 - 6 \gimel_3) + 2 \gimel_1 (5 \gimel_1 - 14 \gimel_2
+ 9 \gimel_3)\rt]\\\nonumber
&&+\frac{4a^2}{3b^2}\kappa \pi G\r\lt[\gimel_0 (27 \gimel_1 - 56 \gimel_2 + 24 \gimel_3) - 2 \gimel_1 (33
\gimel_1 - 74 \gimel_2 + 36 \gimel_3)\rt]
\;.
\ea
We use the results of appendix \ref{appen1} to replace the integrals entering the right side of \eqref{mec8}, with their explicit expressions given in terms of the eccentricity $e$ of the Maclaurin ellipsoid \eqref{bv32}.
It yields
\ba\la{po9l}
\frac{\g_e}{r_e}-\frac{\g_p}{r_p}&=&\frac{6\pi G\r}{e^3}\lt(e - \sqrt{1 - e^2} \arcsin e\rt)-2\pi G\r +
\omega^2+\frac{1}{2c^2}\omega^4a^2\\\nonumber
&&+\frac{\kappa}{6e^7}\lt(-75e - 5e^3 + 122 e^5 - 42 e^7 + 
 3 \sqrt{1 - e^2} (25 + 10 e^2 - 34 e^4 + 8 e^6) \arcsin e\rt)\omega^2\\\nonumber
&&+\frac{\kappa}{24e^9} \pi G\r\lt[525e - 1075 e^3 + 662 e^5 - 112 e^7 - 
 3 \sqrt{1 - e^2} (175 - 300 e^2 + 144 e^4 - 16 e^6) \arcsin e\rt]\lt(E_1+E_2-E_3\rt)\\\nonumber
&&-\frac{3\kappa}{e^5} \pi G\r\lt[3 e (1 - e^2) - \sqrt{1 - e^2} (3 - 2 e^2) \arcsin
e\rt]\lt(E_1-E_2\rt)\\\nonumber
&&-\frac{\kappa}{6e^{9}} \pi G\r(1-e^2)\lt[ e(225  - 351 e^2 - 58 e^4 + 24 e^6) - 
 2  \sqrt{1 - e^2} (225 - 276 e^2 - 85 e^4 + 18 e^6) \arcsin e\rt]\\\nonumber
&&-\frac{\kappa}{2e^{10}} \pi G\r(1-e^2)\lt(75 - 142 e^2 + 38 e^4 + 40 e^6\rt) \arcsin^2 e
\;.
\ea
This form of the Clairaut theorem apparently depends on the gauge parameters $E_1,E_2$ and can be used to impose a constraints on one of them in addition to the constraint given by the level surface equation \eqref{eq1}. For example, we could demand that all post-Newtonian terms in \eqref{po9l} vanish. Such choice of the gauge can be used instead of the constraints \eqref{con1a}, \eqref{con2a} proposed by Bardeen \citep{Bardeen_1971ApJ}.

It is more interesting, however, to bring equation \eqref{po9l} to another form which is more physically relevant, and does not contain explicitly the gauge parameters $E_1, E_2$. This is achieved after
replacement of the eccentricity, $e$, of the Maclaurin ellipsoid with the eccentiricty ${\epsilon}$ of the
PN ellipsoid with the help of the inversion of \eqref{obt5},
\be\la{beg6}
e= {\epsilon}-\kappa\frac{1-{\epsilon}^2}{2{\epsilon}}\lt(E_1-E_2\rt)\;.
\ee
We substitute \eqref{beg6} to \eqref{po9l}, expand in the Taylor series with respect to $\kappa$, and reduce similar terms. This
procedure entirely 
eliminates from \eqref{po9l} the term being proportional to the difference $E_1-E_2$, and yields 
\ba\la{po9s}
\frac{\g_e}{r_e}-\frac{\g_p}{r_p}&=&\frac{6\pi G\r}{{\epsilon}^3}\lt({\epsilon} - \sqrt{1 - {\epsilon}^2}
\arcsin {\epsilon} \rt)-2\pi G\r + \omega^2+\frac{1}{2c^2}\omega^4a^2\\\nonumber
&&+\frac{\kappa}{6e^7}\lt(-75{\epsilon} - 5{\epsilon}^3 + 122 {\epsilon}^5 - 42 {\epsilon}^7 + 
3 \sqrt{1 - {\epsilon}^2} (25 + 10 {\epsilon}^2 - 34 {\epsilon}^4 + 8 {\epsilon}^6) \arcsin
{\epsilon}\rt)\omega^2\\\nonumber
&&+\frac{\kappa}{24{\epsilon}^9} \pi G\r\lt[525{\epsilon} - 1075 {\epsilon}^3 + 662 {\epsilon}^5 - 112
{\epsilon}^7 -
3 \sqrt{1 - {\epsilon}^2} (175 - 300 {\epsilon}^2 + 144 {\epsilon}^4 - 16 {\epsilon}^6) \arcsin
{\epsilon}\rt]\lt(E_1+E_2-E_3\rt)\\\nonumber
&&-\frac{\kappa}{6{\epsilon}^{9}} \pi G\r(1-{\epsilon}^2)\lt[ {\epsilon}(225 - 351 {\epsilon}^2 - 58
{\epsilon}^4 + 24 {\epsilon}^6) -
2 \sqrt{1 - {\epsilon}^2} (225 - 276 {\epsilon}^2 - 85 {\epsilon}^4 + 18 {\epsilon}^6) \arcsin
{\epsilon}\rt]\\\nonumber
&&-\frac{\kappa}{2{\epsilon}^{10}} \pi G\r(1-{\epsilon}^2)\lt(75 - 142 {\epsilon}^2 + 38 {\epsilon}^4 + 40
{\epsilon}^6\rt) \arcsin^2 {\epsilon}
\;.
\ea
The last step is to replace the density $\r$ in \eqref{po9s} with the total mass of the PN ellipsoid by making
use of expression \eqref{vtx4}. We get
\ba\la{pgt5}
\frac{\g_e}{r_e}-\frac{\g_p}{r_p}&=&\frac{3 G
M}{2r_e^2r_p}\frac{3{\epsilon}-{\epsilon}^3-3\sqrt{1-{\epsilon}^2}\arcsin
{\epsilon}}{{\epsilon}^3}+\o^2+\frac{a^2\o^4}{2c^2}\\\nonumber
&-&\frac{{\epsilon} (375 + 25{\epsilon}^2 - 682 {\epsilon}^4 + 
    234{\epsilon}^6) - 
 3 \sqrt{1 -{\epsilon}^2} (125 + 50 {\epsilon}^2 - 194{\epsilon}^4 + 
    40 {\epsilon}^6) \arcsin{\epsilon}}{40 b{\epsilon}^7}\frac{GM}{c^2}\o^2\\\nonumber
    &+&\frac{ {\epsilon} (2625 - 5375 {\epsilon}^2 + 3310 {\epsilon}^4 - 
        704 {\epsilon}^6 + 48 {\epsilon}^8) - 
      \sqrt{
      1 - {\epsilon}^2} (875 -1500 {\epsilon}^2 + 720 {\epsilon}^4 - 
128 {\epsilon}^6) \arcsin{\epsilon}}{640 a^3 b {\epsilon}^9}\frac{3G^2M^2}{c^2}\lt(E_1+E_2-E_3\rt)\\\nonumber            &-&\frac{ {\epsilon} (1-\epsilon^2)(225 -351 {\epsilon}^2 - 58 {\epsilon}^4 + 
                24 {\epsilon}^6) - 
             2  \sqrt{
              1 - {\epsilon}^2} (225 -501 {\epsilon}^2 +191 {\epsilon}^4 + 
175 {\epsilon}^6-42\epsilon^8) \arcsin{\epsilon}}{160 a^3 b {\epsilon}^9}\frac{3G^2M^2}{c^2}\\\nonumber&-&\frac{9}{32}\frac{G^2M^2}{c^2}\frac{75 - 142 {\epsilon}^2 + 38 {\epsilon}^4 +88 {\epsilon}^6
}{a^4{\epsilon}^{10}}\arcsin^2{\epsilon}\;.
\ea
This is the post-Newtonian extension of the classical Clairaut theorem \eqref{clr29}. We can see that the right
side of \eqref{pgt5} depends on the parameters $E_1,E_2,E_3$ solely in the form of the linear combination
$E_1+E_2-E_3$ that is fixed by the physical condition (\ref{eq1}). Like the theorem of Pizetti, the post-Newtonian Clairaut theorem in the form given in \eqref{pgt5} connects the force of gravity on the level surface of the PN ellipsoid with only four parameters -- the post-Newtonian mass $M$, the angular velocity of rotation $\o$, and the equatorial and polar radii, $r_e$ and $r_p$, of the PN ellipsoid.

\section{Practical Applications for Fundamental Astronomy}\la{sec11}

Fundamental astronomy is an essential branch of modern gravitational physics, which explores the fundamental structure of space and time by studying the dynamics of massive bodies and elementary particles, such as photons, in gravitational field on time scales from less than one second to the Hubble time. It establishes basic theoretical principles for high-accuracy calculation and interpretation of various astronomical effects and phenomena observed in gravitationally-bounded systems, for example, clusters of galaxies, the Milky Way, stellar clusters, binary and multiple stars, and the solar system and its sub-systems. It also provides definitions and models that describe astronomical constants, time scales, reference systems and frames used in astronomy and geodesy \citep{Kopeikin_2011_book,Soffel_2013_book}.

Fundamental astronomy obtains physical information on celestial objects and investigates physical laws using the methods of astrometry, celestial mechanics and geodesy which include long baseline radio and optical interferometry, laser and radio ranging, pulsar timing, Doppler tracking, space astrometry, atomic clocks, Global Positioning System (GPS), and other experimental tools like absolute gravimeters, gradientometers, etc. \citep{Soffel_1989_book,brumberg_1991_book,Kopeikin_2014_book1,Kopeikin_2014_book2}. 

We shall apply the formalism of the previous sections to derive practically meaningful post-Newtonian equation of
the level surface and the other relationships used in geodetic and gravimetric applications on the ground and in space. To this end we adopt that the classic geodesy operates with the Maclaurin reference ellipsoid defined by \eqref{bv32} with the semimajor and semimainor axes, $a$ and $b$, and the eccentricity $e$ defined in \eqref{ecc4}. We notice that the Earth oblateness is about $e^2\simeq 1/149=0.0067$ \citep[section
1]{petit_2010} and can be used as a small parameter for expanding all post-Newtonian terms into the convergent Taylor series. When expanding the post-Newtonian formulas to the Taylor series we shall keep the Newtonian expressions as they are, without expansion them with respect to the eccentricity $e$, and take into account only terms of
the order of $e^2$ in the post-Newtonian parts of equations by systematically discarding terms of the order of
${\cal O}\lt(e^4\rt)$, and higher. According to Maclaurin's relation \eqref{pm6y1}, the square of the angular velocity, $\o^2\simeq e^2$, (see
\citep{pizz_1913,chandra_book,Torge_2012_book} for more detail) which allows us to discard terms of the order of ${\cal O}\lt(\o^4\rt)$ and
${\cal O}\lt(\o^2 e^2\rt)$ as well.

\begin{figure}
\begin{center}
\includegraphics[scale=0.5]{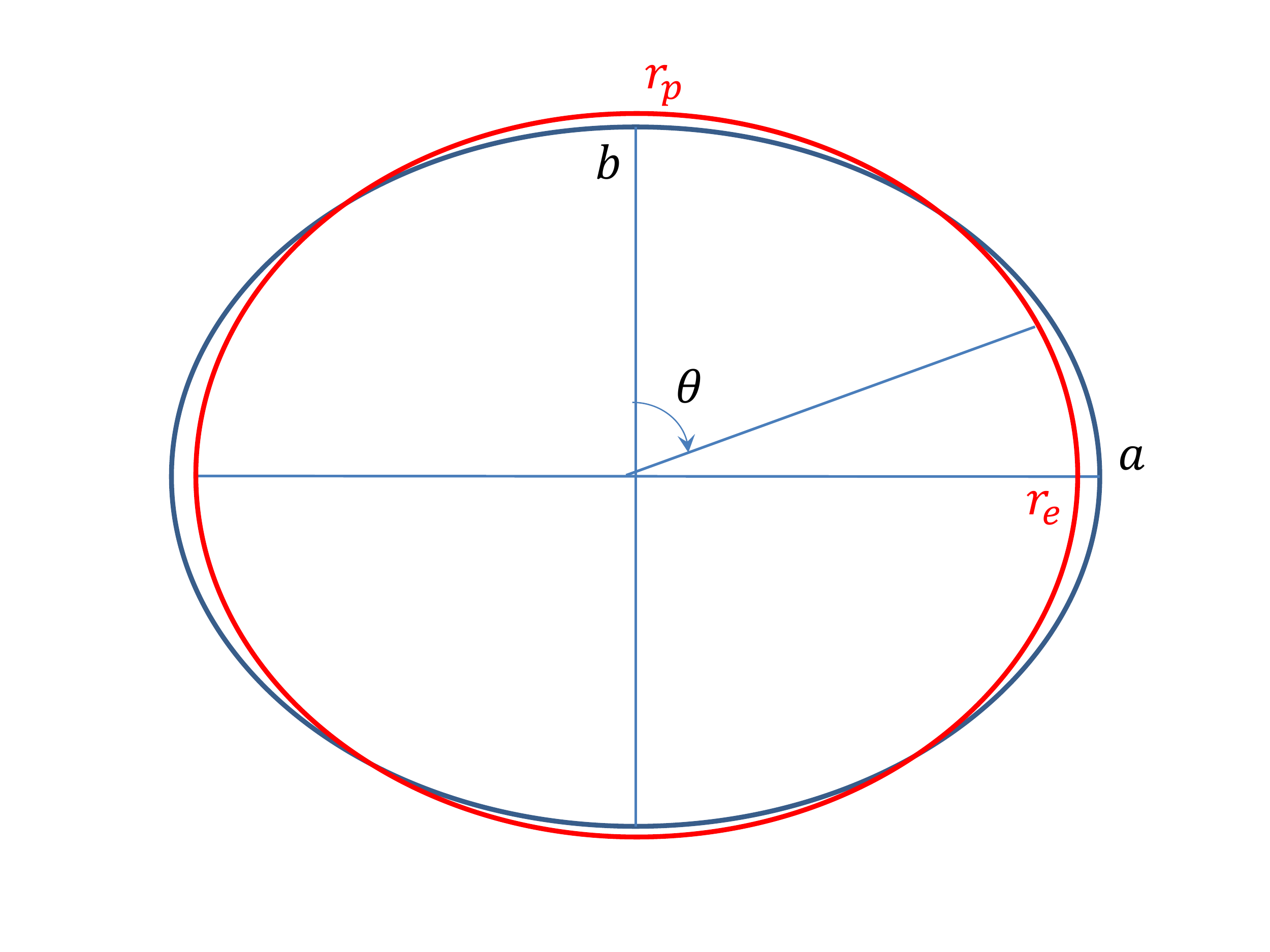}
\caption{Meridional cross-section of the PN ellipsoid (a red curve in the on-line version) versus the Maclaurin ellipsoid (a blue curve in the on-line version) in the Bardeen-Chandrasekhar gauge.
The equatorial, $r_e$, and polar, $r_p$, radii of the PN ellipsoid differ from the semimajor, $a$, and
semiminor, $b$, axes of the Maclaurin ellipsoid, $r_e<a$, $r_p>b$. The maximal radial difference (the
'height' difference) between the surfaces of the PN ellipsoid and the Maclaurin ellipsoid amounts to 1 cm while the difference in the eccentricities $\epsilon$ and $e$ is about $8\times 10^{-7}$. 
\label{fig3}}
\end{center}
\end{figure}

Now, we have to make a decision about what kind of gauge in the mathematical description of the PN ellipsoid would be more preferable for high-precision geodetic applications. First of all, we expand and solve equation \eqref{eq1} of the level surface and find out that the linear combination of the parameters
\be\la{k2x3}
E_1+E_2-E_3=\frac{76}{525}e^4+{\cal O}\lt(e^6\rt)\;,
\ee
that, according to our agreement, can be discarded in all post-Newtonian expressions. Then, we look at the Bardeen gauge conditions \eqref{con1a}, \eqref{con2a} which can be solved with respect to the parameters $E_1$ and $E_2$ with the help of \eqref{k2x3} that is, $E_3=E_1+E_2+{\cal O}\lt(e^4\rt)$. The solution is given by
\be
E_1=-\frac{128}{21}\;,\qquad\qquad E_2=\frac{88}{21}\;,\qquad\qquad E_3=-\frac{40}{21}\;.
\ee
Substituting these values to equation (\ref{1}) we get the equation of the PN ellipsoid in terms of the parameters $a$ and $b$ of the Maclaurin ellipsoid,
\be\la{j7f5}
\frac{\s^2}{a^2}+\frac{z^2}{b^2}=1-\frac{\kappa}{21}\lt(128\frac{\s^4}{a^4}-88\frac{
z^4}{b^4}+40\frac{ \s^2 z^2}{a^2 b^2}\rt)\;,
\ee 
where the numerical value of $\kappa\simeq 5.21\times 10^{-10}$ for the Earth \citep{petit_2010}. Equation \eqref{j7f5} can be reduced to the form
\be\la{vfn7}
\frac{\s^2}{r_e^2}+\frac{z^2}{r_p^2}=1\;,
\ee 
where $r_e$ and $r_p$ are defined in \eqref{eqra} and \eqref{polra},
\be\la{mn3d}
r_e=a\lt(1-\frac{62}{21}\kappa\rt)\;,\qquad\qquad r_p=b\lt(1+\frac{44}{21}\kappa\rt)\;.
\ee
Equation \eqref{vfn7} describes the post-Newtonian deviation from the Maclaurin ellipsoid adopted in the Newtonian-based geodesy. It can be viewed as another ellipsoid with the semimajor axis $r_e$ smaller than $a$ by $\simeq 1.0$ cm, and the semiminor axis $r_p$ larger than $b$ by $\simeq 0.7$ cm (see Fig. \ref{fig3}). The difference \eqref{j7f5} between the PN ellipsoid and the Maclaurin ellipsoid can be also interpreted as a long spatial wave 
\be\la{j7t3w}
\frac{\s^2}{a^2}+\frac{z^2}{b^2}=1-\frac{20}{21}\kappa\lt(1-\frac{27}5\cos 2\theta\rt)\;,
\ee 
with a wavelength equal to a one-half of Earth's radius $R_\oplus$. Bardeen-Chandrasekhar gauge condition lead to a noticeable scale difference between the post-Newtonian and Maclaurin ellipsoids which is not negligibly small for modern geodetic measurements and (in case if the Bardeen-Chandrasekhar gauge is adopted by IERS and IUGG) should be carefully taken into account in the near future adjustment of the geodetic parameters of the reference ellipsoid. 

In our opinion, more preferable would be another choice of the gauge which we shall call the Maclaurin gauge. More specifically, we shall accept that the geodetic coordinate system is chosen in such a way that the shape
parameters $E_1=E_2=0$ exactly. This is always possible due to the residual gauge freedom. In this gauge the
equatorial, $r_e$, and polar, $r_p$, radii are the same as the semimajor and semiminor axes of the Maclaurin
ellipsoid respectively, $r_e=a$, $r_p=b$, and the parameter $E_3=0$ in the above-adopted approximation with all terms of the order of ${\cal O}\lt(e^4\rt)$ having been discarded. The eccentricity of the PN ellipsoid is also equal to that of the
Maclaurin ellipsoid, $\epsilon=e$. With this choice of the residual gauge, the surface of the PN ellipsoid in the given approximation coincides with the surface of the Maclaurin ellipsoid in the Newtonian geodesy,
\be\la{nn25}
\frac{\sigma^2}{a^2}+\frac{z^2}{b^2}=1\;,
\ee
with an error not exceeding $\kappa e^4 R_\oplus\simeq 2\times 10^{-4}$ mm. Such a small error, caused by
the relativistic contribution of Earth's gravitational field, can be safely neglected in all distance/height calculations conducted in the International Terrestrial Reference System (ITRS) defined in \citep{petit_2010}. It also means that the classic
ellipsoidal coordinate system, having been ubiquitously used in the Newtonian geodesy, is not deformed by the relativistic
corrections to the gravity field. 

We can now easily compare the main post-Newtonian equations in the two different gauges.
The Maclaurin relation
\eqref{pm6y1} takes on the following approximate form
\ba\la{nn23}
\frac{\o^2}{2\pi G\r}&=&\frac{\sqrt{1 - e^2} (3 - 2 e^2) \arcsin e-3 e (1 - e^2)}{e^3}+\frac{8}{35}\frac{G
M}{ac^2}e^2\;:\qquad\qquad\,\mbox{the Maclaurin gauge},\\\la{nn23sa}
\frac{\o^2}{2\pi G\r}&=&\frac{\sqrt{1 - \e^2} (3 - 2 \e^2) \arcsin \e-3 \e (1 - \e^2)}{\e^3}+\frac{8}{35}\frac{G
M}{ac^2}e^2\;:\qquad\qquad\,\mbox{the Bardeen-Chandrasekhar gauge}\;.
\ea
which, after replacing $\r$ with the help of the inverse of \eqref{totm0} and expansion, can be recast to
\ba\la{mk9fs}
\o^2&=&\frac{3 G M}{2a^3}\lt[\frac{ (3 - 2 e^2) \arcsin e-3 e \sqrt{1 -
e^2}}{e^3}-\frac{4}{7}\frac{GM}{ac^2}e^2\rt]\;:\qquad\quad\,\mbox{the Maclaurin gauge},\\\la{mk5sf}
\o^2&=&\frac{3 G M}{2r_e^3}\lt[\frac{ (3 - 2 \e^2) \arcsin \e-3 \e \sqrt{1 -
\e^2}}{\e^3}-\frac{4}{7}\frac{GM}{ac^2}e^2\rt]\;:\qquad\quad\,\mbox{the Bardeen-Chandrasekhar gauge}\;.
\ea
Numerical values for the geopotential and the semimajor axis of reference ellipsoid are given in \citep[Table 1.1]{petit_2010}. They yield, $GM_\oplus/c^2 R_\oplus \simeq 6.7\times 10^{-10}$. The magnitude of the relativistic correction to the classic Maclaurin relation between the angular velocity $\o$ of the rotating ellipsoid and its oblateness $e$ are given in the Bardeen-Chandrasekhar gauge by the post-Newtonian terms that appear as a consequence of the expansion of the terms with $\e$ in the right side of \eqref{mk5sf}. These post-Newtonian terms amount to  $1.3\times 10^{-10}$. On the other hand, the relativistic correction to $\o$ in the Maclaurin gauge is given solely by the last term in the right part of  \eqref{mk9fs}. Its magnitude is reduced by the factor of $e^2$ and amounts to only $2.6\times 10^{-12}$. 

The approximate version of the post-Newtonian theorem of Pizzetti \eqref{pipi1} reads
\ba\la{nnn3}
2\frac{\g_e}{a}+\frac{\g_p}{b}&=&- 2\o^2+ \frac{3 G M}{a^2b} +\frac{4G M}{ac^2}
\lt[\lt(3+\frac{47}{25}e^2\rt)\frac{GM}{a^3}+\frac14\o^2\rt]\;:\qquad\quad\quad\mbox{the Maclaurin gauge},\\\la{swu7}
2\frac{\g_e}{r_e}+\frac{\g_p}{r_p}&=&- 2\o^2+ \frac{3 G M}{r_e^2r_p} +\frac{4G M}{ac^2}
\lt[\lt(3+\frac{47}{25}e^2\rt)\frac{GM}{a^3}+\frac14\o^2\rt]\;:\qquad\qquad\mbox{the Bardeen-Chandrasekhar gauge}\;,
     \ea
where the relation of the radii $r_e$ and $r_p$ to $a$ and $b$ respectively, are given in \eqref{mn3d}. 
The post-Newtonian corrections in the Clairaut theorem \eqref{pgt5}, after it is expanded with respect to the eccentricity $e$, yield
\begin{flalign}
\la{nnn35}
\frac{\g_e}{a}-\frac{\g_p}{b}&=&\o^2+\frac{3 G M}{2a^2b}\frac{e(3-e^2)-3\sqrt{1-e^2}\arcsin
e}{e^3}+\frac{GM}{ac^2}\lt(\frac{59}{25} \frac{G M}{a^3}e^2 +
   \frac{11}{14} \o^2\rt)\,:\,\qquad\qquad\quad\;\;\mbox{the Maclaurin gauge},\\\la{ssr5}
   \frac{\g_e}{r_e}-\frac{\g_p}{r_p}&=&\o^2+\frac{3 G M}{2r_e^2r_p}\frac{e(3-e^2)-3\sqrt{1-e^2}\arcsin
   e}{e^3}+\frac{GM}{ac^2}\lt(\frac{59}{25} \frac{G M}{a^3}e^2 +
      \frac{11}{14} \o^2\rt)\,:\,\mbox{the Bardeen-Chandrasekhar gauge}.
   \end{flalign}
The post-Newtonian corrections to the gravitational field entering the the Pizzetti and Clairaut theorems \eqref{nnn3}--\eqref{ssr5} are not so negligibly small, amount to the magnitude of approximately 3 $\mu$Gal (1 Gal = 1 cm/s$^2$), and are to be taken into account in calculation of the parameters of the reference-ellipsoid from astronomical and gravimetric data in a foreseeable future.

The Bardeen-Chandrasekhar gauge has some advantage in comparison of the masses of the Newtonian and post-Newtonian ellipsoids. According to the Bardeen-Chandrasekhar gauge condition the masses of the two ellipsoids are exactly the same, $M=M_{\rm Maclaurin}$. On the other hand, if we chose the Maclaurin gauge, the masses of the two ellipsoids will differ. Post-Newtonian correction to the Newtonian mass of the Earth, $M_{\rm Maclaurin}=M_N=4\pi\r a^2b/3$, can be evaluated in the Maclaurin gauge from \eqref{totm1a} which is given in the approximation under consideration, by
\be\la{mn5gcx}
M=M_N\lt[1+2\kappa\lt({\gimel}_0+\frac{\o^2}{5\pi G\r}\rt)\rt]\;.
\ee
Because the mass couples with the universal gravitational constant $G$, it contributes to the numerical value of the geocentric gravitational constant $GM_\oplus=3.986004418\times 10^{14}$ m$^3$s$^{-2}$ \citep[Table 1.1]{petit_2010}. After expansion of the right side of \eqref{mn5gcx} with respect to the eccentricity $e$, the relativistic variation in the value of $GM_\oplus$ is 
\be\la{tnuf5x}
\frac{\delta (GM_\oplus)_{pN}}{GM_\oplus}\simeq 2\kappa{\gimel}_0\simeq 4{\kappa}\simeq 2.8\times 10^{-9} \;.
\ee
The current uncertainty in the numerical value of $GM_\oplus$ is $8\times 10^5$ m$^3$s$^{-2}$ \citep[Table 1.1]{petit_2010} which gives the fractional uncertainty
\be\la{zc32}
\frac{\delta (GM_\oplus)}{GM_\oplus}\simeq 2.0\times 10^{-9}\;.
\ee
This is comparable with the relativistic contribution \eqref{tnuf5x} which must be taken into account in the reduction of precise geodetic data processing if the Maclaurin gauge is adopted in the post-Newtonian geodesy. 

Similar considerations tells us that in the Bardeen-Chandrasekhar gauge the Newtonian and post-Newtonian values of the angular momentum of the Earth are exactly the same, $S=S_{\rm Maclaurin}$. However, if we chose the Maclaurin gauge we should expect the difference between the Newtonian and relativistic angular momenta given by \eqref{angu1} which, in the approximation where all terms of the order of ${\cal O}\lt(e^4\rt)$ are discarded, reads
\be\la{nevxy} 
S=S_N\lt[1+\frac27\kappa\lt(12\gimel_0-16\gimel_1+\frac{3\o^2}{5\pi G\r}\rt)\rt]\;.
\ee
The fractional difference between the relativistic and Newtonian angular momenta of the Earth is
\be\la{ersvxu}
\frac{\delta S}{S}\simeq\frac{40}{7}\gimel_1\simeq 2\times 10^{-9}\;.
\ee
It should be noted however that the difference \eqref{ersvxu} is not so important in geodesy because the angular momentum of the earth is not yet directly measurable quantity as contrasted with the angular velocity $\o$ which is measured directly by Very Long Baseline Interferometry (VLBI) of the IERS. 
\newline\newline
\noindent
{\bf Asknowledgements}
\newline
\noindent
We thank anonymous referees for their valuable comments and suggestions for improving the presentation of the manuscript.
The work of S.~M. Kopeikin and E.~M. Mazurova has been supported by the grant \textnumero 14-27-00068 of the
Russian Scientific Foundation. The research of W.-B. Han has been funded by the National Natural Science Foundation
of China (grant \textnumero 11273045) and the China Scholarship Council Fellowship \textnumero 201304910030. E.~M. Mazurova and W.-B. Han are grateful to the Department of Physics and Astronomy of the University of Missouri in Columbia (USA) for hospitality.

\bibliographystyle{unsrt}
\bibliography{PN_spheroid_bib}

\begin{thebibliography}{10}
\bibitem{footnote-1}
{Speaking more rigorously, the surface of the fourth order is to be called a
  spheroid. We shall show in the last section \ref{sec11} of this paper that,
  by making use of the coordinate freedom of general relativity, the deviations
  of the shape of the PN-ellipsoid from the Maclaurin ellipsoid can be made so
  small that they can be ignored in practical applications.}
\end{thebibliography}

\appendix
\section{Integrals}\la{appen1}

Integrals entering equations \eqref{11aa}, \eqref{11bb}, \eqref{wdv6}, \eqref{m95}, \eqref{i1as}, \eqref{i1an}
are solved in two steps:
\begin{enumerate} 
\item integrating with respect to the angle $\lambda$ from $0$ to $2\pi$, 
\item integrating with respect to the angle $\th$ from $0$ to $\pi$ by making use of a new variable
\begin{equation}\label{13}
u=b^2\tan^2\th\;,\qquad du=2b^2\sec^3\th \sin\th d\th\;, 
\end{equation}
which changes from $0$ to $\infty$.
\end{enumerate}
In terms of the new variable we have
\be\la{14}
\sin\th=\sqrt{\frac{u}{b^2+u}}\;,\qquad\cos\th=\frac{b}{\sqrt{b^2+u}}\;,
\ee
and
\be\la{14as}
A=\frac1{a^2}\frac{a^2+u}{b^2+u}\;,\qquad \frac{d\Omega}{A}=\frac{a^2b}{2}\frac{d\l du}{(a^2+u)\sqrt{b^2+u}}\;.\ee
We use these values for transforming integrands in \eqref{11aa}, \eqref{11bb} where we also take into account
that all functions entering the integrands are even functions of the argument, $f(\cos\th)=f(-\cos\th)$. Thus,
it makes \be\int_0^\pi f(\cos\th)\sin\th d\th=2\int_0^{\pi/2} f(\cos\th)\sin\th d\th\;.\ee.
This procedure allows us to represent the integrals under discussion in the following form:
\allowdisplaybreaks
\begin{subequations}
\ba
&&{\cal J}_0\equiv\oint_{S^2}\frac{d\Omega}{A}=2\pi a^2b\int\limits_0^{\infty}\frac{du}{(a^2 + u) \sqrt{b^2 +
u}}\;,\\
&&{\cal J}_1\equiv\oint_{S^2}\frac{\cos^2\th}{A^2}d\Omega=2\pi a^4 b^3\int\limits_0^{\infty}\frac{du}{(a^2 +
u)^2 \sqrt{b^2 + u}}\;,\\
&&{\cal J}_2\equiv\oint_{S^2}\frac{ B}{A^2}\cos\th d\Omega=2\pi a^4 b z\int\limits_0^{\infty}\frac{du}{(a^2 +
u)^2 \sqrt{b^2 + u}}\;,\\
&&{\cal J}_3\equiv\oint_{S^2}\frac{B}{A^3}\cos^3\th d\Omega=2\pi a^6b^3 z\int\limits_0^{\infty}\frac{du}{(a^2 +
u)^3 \sqrt{b^2 + u}}\;,\\
&&{\cal J}_4\equiv\oint_{S^2}\frac{\cos^4\th}{A^3}d\Omega=2\pi a^6b^5\int\limits_0^{\infty}\frac{du}{(a^2 +
u)^3 \sqrt{b^2 + u}}\;,\\
&&{\cal J}_5\equiv\oint_{S^2}\frac{B^2}{A^2}d\Omega=\frac{\pi a^2}{b}\lt[\int\limits_0^{\infty}\frac{b^2 u + (2
a^2 - u) z^2}{(a^2 + u)^2 \sqrt{b^2 + u}}du+
b^2 C({\bm x})\int\limits_0^{\infty}\frac{udu}{(a^2 + u)^2 \sqrt{b^2 + u}}\rt]\;,\\
&&{\cal J}_6\equiv\oint_{S^2}\frac{B^2}{A^4}\cos^4\th d\Omega=\pi a^6b^3\lt[\int\limits_0^{\infty}\frac{b^2 u +
(2 a^2 - u) z^2}{(a^2 + u)^4 \sqrt{b^2 + u}}du
+b^2C({\bm x})\int\limits_0^{\infty}\frac{udu}{(a^2 + u)^4 \sqrt{b^2 + u}}\rt]\;,\\
&&{\cal J}_7\equiv\oint_{S^2}\frac{B^2}{A^3}\cos^2\th d\Omega=\pi a^4b\lt[\int\limits_0^{\infty}\frac{b^2 u +
(2 a^2 - u) z^2}{(a^2 + u)^3 \sqrt{b^2 + u}}du+
b^2 C({\bm x})\int\limits_0^{\infty}\frac{ udu}{(a^2 + u)^3 \sqrt{b^2 + u}}\rt]\;,\\
&&{\cal J}_8\equiv\oint_{S^2}\frac{B^3}{A^4}\cos^3\th d\Omega=\pi a^6b z\lt[\int\limits_0^{\infty}\frac{3b^2 u
+ (2 a^2 - 3u) z^2}{(a^2 + u)^4 \sqrt{b^2 + u}}du
+3b^2C({\bm x})\int\limits_0^{\infty}\frac{udu}{(a^2 + u)^4 \sqrt{b^2 + u}}\rt]\;,\\
&&{\cal J}_9\equiv\oint_{S^2}\frac{B^4}{A^5}\cos^4\th d\Omega=\frac{\pi
a^6b}{4}\int\limits_0^{\infty}\frac{3b^4 u^2+6 b^2u (4 a^2 - u) z^2 + (8 a^4 - 24 a^2 u + 3 u^2) z^4}{(a^2 +
u)^5 \sqrt{b^2 + u}}du\\\nonumber
&&\phantom{++++++}+\frac32 C({\bm x})\pi a^6b^3\lt[\int\limits_0^{\infty}\frac{\lt(b^2-z^2\rt)u^2+4 a^2 u
z^2}{(a^2 + u)^5 \sqrt{b^2 + u}}du
+\frac{b^2}2 C({\bm x})\int\limits_0^{\infty}\frac{u^2du}{(a^2 + u)^5 \sqrt{b^2 + u}}\rt]\;,\\
&&{\cal J}_{10}\equiv\oint_{S^2}\frac{B^3}{A^3}\sin\th\cos\l d\Omega=\frac{3\pi a^2
x}{4b}\lt[\int\limits_0^{\infty}\frac{\lt(b^2 u + 4 a^2 z^2 - u z^2\rt)udu}{\lt(a^2 +
u\rt)^3\sqrt{b^2+u}}+b^2C({\bm x})\int\limits_0^{\infty}\frac{u^2du}{\lt(a^2 + u\rt)^3\sqrt{b^2+u}}\rt]\;, \\
&&{\cal J}_{11}\equiv\oint_{S^2}\frac{B^3}{A^3}\sin\th\sin\l d\Omega=\frac{3\pi a^2
y}{4b}\lt[\int\limits_0^{\infty}\frac{\lt(b^2 u + 4 a^2 z^2 - u z^2\rt)udu}{\lt(a^2 +
u\rt)^3\sqrt{b^2+u}}+b^2C({\bm x})\int\limits_0^{\infty}\frac{u^2du}{\lt(a^2 + u\rt)^3\sqrt{b^2+u}}\rt]\;, \\
&&{\cal J}_{12}\equiv\oint_{S^2}\frac{B^3}{A^3}\cos\th d\Omega=\frac{\pi a^4
z}{b}\lt[\int\limits_0^{\infty}\frac{\lt(3b^2 u +2 a^2 z^2 -3 u z^2\rt)du}{\lt(a^2 +
u\rt)^3\sqrt{b^2+u}}+3b^2C({\bm x})\int\limits_0^{\infty}\frac{udu}{\lt(a^2 + u\rt)^3\sqrt{b^2+u}}\rt]\;,\\
&&{\cal J}_{13}\equiv\oint_{S^2}\frac{B}{A^2}\sin\th\cos\l d\Omega =\pi a^2 b
x\int\limits_0^{\infty}\frac{udu}{\lt(a^2 + u\rt)^2\sqrt{b^2+u}}\;,\\
&&{\cal J}_{14}\equiv\oint_{S^2}\frac{B}{A^2}\sin\th\sin\l d\Omega =\pi a^2 b
y\int\limits_0^{\infty}\frac{udu}{\lt(a^2 + u\rt)^2\sqrt{b^2+u}}\;,\\
&&{\cal J}_{15}\equiv\oint_{S^2}\frac{B^4}{A^3} d\Omega=\frac{\pi a^2}{4b^3}\int\limits_0^{\infty}\frac{3b^4
u^2+6 b^2u (4 a^2 - u) z^2 + (8 a^4 - 24 a^2 u + 3 u^2) z^4}{(a^2 + u)^3 \sqrt{b^2 + u}}du\\\nonumber
&&\phantom{++++++}+\frac32 C({\bm x})\frac{\pi a^2}{b}\lt[\int\limits_0^{\infty}\frac{\lt(b^2-z^2\rt)u^2+4 a^2
u z^2}{(a^2 + u)^3 \sqrt{b^2 + u}}du
+\frac{b^2}2 C({\bm x})\int\limits_0^{\infty}\frac{u^2du}{(a^2 + u)^3 \sqrt{b^2 + u}}\rt]\;,\\
&&{\cal J}_{16}\equiv\oint_{S^2}\frac{B^4}{A^4}\cos^2\th d\Omega=\frac{\pi
a^4}{4b}\int\limits_0^{\infty}\frac{3b^4 u^2+6 b^2u (4 a^2 - u) z^2 + (8 a^4 - 24 a^2 u + 3 u^2) z^4}{(a^2 +
u)^4 \sqrt{b^2 + u}}du\\\nonumber
&&\phantom{++++++}+\frac32 C({\bm x})\pi a^4b\lt[\int\limits_0^{\infty}\frac{\lt(b^2-z^2\rt)u^2+4 a^2 u
z^2}{(a^2 + u)^4 \sqrt{b^2 + u}}du
+\frac{b^2}2 C({\bm x})\int\limits_0^{\infty}\frac{u^2du}{(a^2 + u)^4 \sqrt{b^2 + u}}\rt]\;,
\ea
\end{subequations}
These integrals can be performed analytically,
\begin{subequations}
\ba
{\cal J}_0&\equiv&2\pi a^2 {\gimel}_0\;,\\
{\cal J}_1&\equiv&2\pi a^2 b^2{\gimel}_1\;,\\
{\cal J}_2&=&2\pi a^2 z{\gimel}_1\;,\\
{\cal J}_3&\equiv&2\pi a^2b^2 z{\gimel}_2\;,\\
{\cal J}_4&\equiv&2\pi a^2b^4{\gimel}_2\;,\\
{\cal J}_5&=&\pi a^2\lt[\lt(1 - \frac{z^2}{b^2}\rt){\gimel}_0- \lt(1 - 3\frac{z^2}{b^2}\rt){\gimel}_1+C({\bm
x})\lt({\gimel}_0-{\gimel}_1\rt)\rt]\;,\\
{\cal J}_6&\equiv&\pi a^2b^4\lt[\lt(1-\frac{z^2}{b^2}\rt){\gimel}_2-\lt(1-\frac{3z^2}{b^2}\rt){\gimel}_3
+C({\bm x})\lt({\gimel}_2-{\gimel}_3\rt)\rt]\;,\\
{\cal J}_7&=&\pi a^2b^2\lt[\lt(1 - \frac{z^2}{b^2}\rt){\gimel}_1-\lt(1 -
3\frac{z^2}{b^2}\rt){\gimel}_{2}+C({\bm x})\lt({\gimel}_1-{\gimel}_2\rt)\rt]\;,\\
{\cal J}_8&=&3\pi a^2b^2 z\lt[\lt(1 - \frac{z^2}{b^2}\rt){\gimel}_{2}-\lt(1 -
\frac53\frac{z^2}{b^2}\rt){\gimel}_{3} +C({\bm x})\lt({\gimel}_2-{\gimel}_3\rt)\rt]\;,\\
{\cal J}_9&=&\frac{3}{4}\pi a^2b^4\lt[\lt(1 - \frac{z^2}{b^2}\rt)^2{\gimel}_{2}-2\lt(1 -
6\frac{z^2}{b^2}+5\frac{z^4}{b^4}\rt){\gimel}_{3}+\lt(1 -
10\frac{z^2}{b^2}+\frac{35}{3}\frac{z^4}{b^4}\rt){\gimel}_{4} \rt]\\\nonumber
&+&\frac32 C({\bm x})\pi
a^2b^4\lt[\lt(1-\frac{z^2}{b^2}\rt){\gimel}_2-2\lt(1-\frac{3z^2}{b^2}\rt){\gimel}_3+\lt(1-\frac{5z^2}{b^2}\rt){\gimel}_4\rt]\\\nonumber&+&\frac34
C^2({\bm x})\pi a^2b^4\lt({\gimel}_2-2{\gimel}_3+{\gimel}_4\rt)
\;,\\
{\cal J}_{10}&=&\frac34\pi a^2
x\lt[\lt(1-\frac{z^2}{b^2}\rt)\gimel_0-2\lt(1-\frac{3z^2}{b^2}\rt)\gimel_1+\lt(1-\frac{5z^2}{b^2}\rt)\gimel_2+C({\bm
x})\lt(\gimel_0-2\gimel_1+\gimel_2\rt)\rt] \;,\\
{\cal J}_{11}&=& \frac34\pi a^2
y\lt[\lt(1-\frac{z^2}{b^2}\rt)\gimel_0-2\lt(1-\frac{3z^2}{b^2}\rt)\gimel_1+\lt(1-\frac{5z^2}{b^2}\rt)\gimel_2++C({\bm
x})\lt(\gimel_0-2\gimel_1+\gimel_2\rt)\rt] \;,\\
{\cal J}_{12}&=& 3\pi a^2
x\lt[\lt(1-\frac{z^2}{b^2}\rt)\gimel_1-\lt(1-\frac53\frac{z^2}{b^2}\rt)\gimel_2+C({\bm
x})\lt(\gimel_1-\gimel_2\rt)\rt] \;,\\
{\cal J}_{13}&=&\pi a^2 x\lt(\gimel_0-\gimel_1 \rt)\;,\\
{\cal J}_{14}&=&\pi a^2 y\lt(\gimel_0-\gimel_1 \rt)\;,\\
{\cal J}_{15}&=&\frac{3}{4}\pi a^2\lt[\lt(1 - \frac{z^2}{b^2}\rt)^2{\gimel}_{0}-2\lt(1 -
6\frac{z^2}{b^2}+5\frac{z^4}{b^4}\rt){\gimel}_{1}+\lt(1 -
10\frac{z^2}{b^2}+\frac{35}{3}\frac{z^4}{b^4}\rt){\gimel}_{2} \rt]\\\nonumber
&+&\frac32 C({\bm x})\pi a^2
\lt[\lt(1-\frac{z^2}{b^2}\rt){\gimel}_0-2\lt(1-\frac{3z^2}{b^2}\rt){\gimel}_1+\lt(1-\frac{5z^2}{b^2}\rt){\gimel}_2\rt]\\\nonumber&+&\frac34
C^2({\bm x})\pi a^2\lt({\gimel}_0-2{\gimel}_1+{\gimel}_2\rt)
\;,\\
{\cal J}_{16}&=&\frac{3}{4}\pi a^2b^2\lt[\lt(1 - \frac{z^2}{b^2}\rt)^2{\gimel}_{1}-2\lt(1 -
6\frac{z^2}{b^2}+5\frac{z^4}{b^4}\rt){\gimel}_{2}+\lt(1 -
10\frac{z^2}{b^2}+\frac{35}{3}\frac{z^4}{b^4}\rt){\gimel}_{3} \rt]\\\nonumber
&+&\frac32 C({\bm x})\pi a^2b^2
\lt[\lt(1-\frac{z^2}{b^2}\rt){\gimel}_1-2\lt(1-\frac{3z^2}{b^2}\rt){\gimel}_2+\lt(1-\frac{5z^2}{b^2}\rt){\gimel}_3\rt]\\\nonumber&+&\frac34
C^2({\bm x})\pi a^2\lt({\gimel}_1-2{\gimel}_2+{\gimel}_3\rt)
\;,
\ea
\end{subequations}
where 
\be
{\gimel}_n\equiv
a^{2n}b\int\limits_0^\infty\frac{du}{(a^2+u)^{n+1}\sqrt{b^2+u}}=\frac{2\sqrt{1-e^2}}{e^{2(n+1)}}\int\limits^\infty_{\frac{\sqrt{1-e^2}}{e}}\frac{d\xi}{\lt(1+\xi^2\rt)^{n+1}}\;.
\ee
is a table integral given in (Gradshteyn and Ryzhik, integral 2.148-4)
\be
{\gimel}_n=\frac{(2n-1)!!}{2^{n-1}n!}\lt(\frac{\sqrt{1-e^2}}{e^{2n+1}}\arcsin
e-\lt(1-e^2\rt)\sum\limits_{k=1}^{n}\frac{(n-k)!}{(2n-2k+1)!!}\frac{2^{n-k}}{e^{2k}}\rt)
\;.
\ee
In particular,
\begin{subequations}
\ba\la{inta0}
{\gimel}_0&=&2\frac{\sqrt{1 - e^2}}{e} \arcsin e \;,
\\\la{inta1}
{\gimel}_1&=&\frac{\sqrt{1-e^2}}{e^3}\arcsin e-\frac{1-e^2}{e^2}\;,
\\\la{inta2}
{\gimel}_2&=&\frac{3}{4}\frac{\sqrt{1-e^2}}{e^5}\arcsin e-\frac{(1-e^2)(3+2e^2)}{4e^4}\;,
\\
{\gimel}_3&=&\frac{5}{8}\frac{\sqrt{1-e^2}}{e^7}\arcsin e-\frac{(1-e^2)(15+10e^2+8e^4)}{24e^6}\;,
\\
{\gimel}_4&=&\frac{35}{64}\frac{\sqrt{1-e^2}}{e^9}\arcsin e-\frac{(1-e^2)(105+70e^2+56e^4+48e^6)}{192e^8}
\ea
\end{subequations}
There is also a recurrent formula that allows us to calculate the integrals ${\gimel}_n$ by iterations starting
from $\gimel_0$,
\be
{\gimel}_n=\frac{2n-1}{2n}\frac{{\gimel}_{n-1}}{e^2}-\frac{1-e^2}{ne^2}\qquad\qquad(n\ge 1)\;.
\ee
where $\gimel_0$ is given in \eqref{inta0}. Notice that none of these integrals is divergent for small values
of the eccentricity because the denominators of the integrals suppress the small value of the eccentricity in
the numerator.

\end{document}